\documentclass[10pt,aps,prapplied,twocolumn,superscriptaddress,longbibliography,floatfix]{revtex4-2}

\usepackage[utf8]{inputenc}
\usepackage[T1]{fontenc}
\usepackage{amsmath,amssymb,mathtools}
\usepackage{graphicx}
\usepackage{placeins}
\usepackage{xcolor}
\usepackage{xurl}
\usepackage{hyperref}
\usepackage{times}
\usepackage{orcidlink}
\usepackage{upgreek}
\hypersetup{colorlinks=true,linkcolor=blue!70!black,citecolor=blue!70!black,urlcolor=blue!70!black}

\graphicspath{{figures/}}

\newcommand{\nzz}{n_{ZZ}}
\newcommand{\ket}[1]{\left|#1\right\rangle}
\newcommand{\nB}{n_B}
\newcommand{\nq}{n_q}
\newcommand{\RObservedVariance}{6.150\times10^{-3}}
\newcommand{\RMeanEstimationVariance}{2.070\times10^{-4}}
\newcommand{\RExcessVariance}{5.943\times10^{-3}}
\newcommand{\RVarianceRatio}{29.7}

\newcommand{\CObservedVariance}{4.260\times10^{-6}}
\newcommand{\CMeanEstimationVariance}{1.776\times10^{-4}}

\newcommand{\CVariancePercent}{2.4}

\newcommand{\RInflationThreshold}{31.5}

\newcommand{\ControlBiasReductionMin}{35.2}
\newcommand{\ControlBiasReductionMax}{50.5}
\newcommand{\ControlProductReduction}{41.3}

\newcommand{\BlockLargeWeight}{74.4}
\newcommand{\BlockFiveWeight}{3.7}
\newcommand{\BlockMaxQuantile}{3.1295}
\newcommand{\BlockFiveDefaults}{0.183}

\newcommand{\BlockFiveMatched}{0.234}

\newcommand{\BlockFiveEOne}{0.116}
\newcommand{\BlockFiveEOneLow}{-0.578}
\newcommand{\BlockFiveEOneHigh}{-0.197}
\newcommand{\BlockFiveETwo}{0.285}
\newcommand{\BlockFiveETwoLow}{-0.403}
\newcommand{\BlockFiveETwoHigh}{-0.028}
\newcommand{\BlockFiveEThree}{0.230}
\newcommand{\BlockFiveEThreeLow}{-0.471}
\newcommand{\BlockFiveEThreeHigh}{-0.069}

\newcommand{\ReadoutCOne}{0.9549}
\newcommand{\ReadoutCTwo}{0.9616}
\newcommand{\ReadoutCThree}{0.9715}
\newcommand{\ReadoutFractionMin}{3.7}
\newcommand{\ReadoutFractionMax}{5.4}

\newcommand{\nt}{n_t} 
\newcommand{\kopt}{k_{\mathrm{opt}}}
\newcommand{\Fest}{\mathcal{F}_{\mathrm{est}}}

\newcommand{\Phw}{P_{\mathrm{hw}}}

\begin{document}

\title{Classical simulability and hardware diagnostics for QAOA\\
on geographic coverage benchmarks}

\author{F.~S.~Luiz\orcidlink{0000-0002-6375-0939}}%\footnote[1]{$^{*}$Corresponding author
\affiliation{Instituto de F\'isica Gleb Wataghin, UNICAMP, Campinas, SP, Brazil}
\author{D.~Abreu}
\affiliation{Universidade Federal do Par\'a (UFPA), Bel\'em, PA, Brazil}
\author{A.~J.~G.~Abel\'em}
\affiliation{Universidade Federal do Par\'a (UFPA), Bel\'em, PA, Brazil}
\author{J.~E.~da~Fonseca}
\affiliation{Venturus Centro de Inovação Tecnol\'ogica, Campinas, SP, Brazil}
\author{A.~F.~P.~Santos}
\affiliation{Instituto Militar de Engenharia, Rio de Janeiro, RJ, Brazil}
\author{M.~C.~de~Oliveira\orcidlink{0000-0003-2251-2632}}
\affiliation{Instituto de F\'isica Gleb Wataghin, UNICAMP, Campinas, SP, Brazil}

\date{November 25, 2026}

\begin{abstract}
Interpreting a quantum-optimization experiment requires checking what the circuit encodes and what its measured observables can resolve. We demonstrate a three-step diagnostic procedure on static geographic coverage benchmarks: validate the encoding, compute an ideal structural reference, and compare measured outputs with that reference at the available shot budget. In the 24-qubit circuit, 144 infeasible ground states lie below 288 minimum covers; the ideal circuit factorizes into 13 blocks of at most four qubits. A faithful 51-qubit slack encoding preserves the block partition and provides an exact reference for larger blocks. At its executed angles, the ideal optimum probability is $5.47\times10^{-20}$, so zero optimum counts in $10^5$ shots cannot diagnose hardware loss. Energy alone also obscures the output structure: a product distribution with measured bit marginals accounts for about $90\%$ of the observed reduction. In three fixed-circuit jobs, a global weighted-covariance ratio near $0.50$ conceals a five-variable block with ratios $0.12$--$0.28$ and only $\BlockFiveWeight\%$ of the ideal global weight. Its deficit relative to the aggregate is resolved by simultaneous uncertainty bands conditional on the observed counts. The procedure identifies which circuit-level discrepancies are measurable, while leaving their physical cause and device stability undetermined.
\end{abstract}

\maketitle

\section{Introduction}
\label{sec:intro}
Quantum-optimization experiments require a reference that separates the encoded task, the chosen circuit parameters and the effects of execution. For QAOA \cite{farhi2014quantum,zhou2020qaoa}, register size and energy reduction alone do not supply that reference. Experiments on superconducting processors and analyses of routing and noise illustrate the need to account for the executed circuit and its classical alternatives \cite{harrigan2021nonplanar,weidenfeller2022scaling,franca2021limitations}.

We use static geographic coverage instances constructed from surveyed Pelot\~oes Especiais de Fronteira (PEF) sites \cite{MinistDefesa2023} and the official IBGE 2024 municipality product for the Brazilian Legal Amazon \cite{IBGE2024AmazoniaLegal}. The task is minimum set cover when bases and targets are distinct and minimum dominating set when municipalities serve as both. The former is the location set covering problem \cite{toregas1971location}, distinct from fixed-budget maximal coverage \cite{church1974maximal}. These problems are NP-hard, including relevant unit-disk settings, while geometric structure permits approximation schemes \cite{karp1972reducibility,vazirani2001approximation,clark1990unit,fowler1981optimal,johnson1982np,hunt1998nc,nieberg2006ptas,mustafa2010improved}. Our finite instances have certified classical optima. Coverage means that a target lies within a prescribed radius of an active base; the model does not include vehicle routes or operational drone surveillance.

Graph decomposition and tensor-network simulation provide established references for variational circuits \cite{markov2008simulating,medvidovic2021classical,dupont2022calibrating,lykov2023sampling}. Ponce \emph{et al.}\ decompose combinatorial problems before quantum execution \cite{ponce2025graph}. Here we retain the submitted circuit and use its support components to calculate exact ideal observables. Cogburn \emph{et al.}\ compare hardware with a noiseless finite-shot matrix-product-state reference under the same shallow ansatz and optimizer budget \cite{cogburn2026honeypot}. Our reference conditions on the executed angles and requires no optimizer replay. Sharma and Lau use an independent-error gate-count proxy to contextualize compiled optimization circuits \cite{sharma2026circuits}; we distinguish such resource proxies from measured output moments. LR-QAOA benchmarks compare objective performance with random sampling across width and depth \cite{montanez2025evaluating}; our measured-marginal and block references instead ask which correlations are needed to explain a given energy reduction.

The contribution is a diagnostic procedure connecting encoding validation, an exact circuit reference and observable selection. It exposes three interpretation failures in the executed benchmarks: infeasible surrogate minima can be favored over minimum covers, zero optimum counts can be expected even without hardware error, and aggregate energy and covariance can conceal a local discrepancy. In the faithful circuit, a five-variable block contributes only $\BlockFiveWeight\%$ of the ideal global weighted covariance yet lies below the global retention ratio in all five observed contexts under the stated conditional counting model. The structural identities are established tools; their use here determines which comparisons the experiment can resolve. Detecting and localizing a physical error process requires additional evidence, as in dedicated crosstalk protocols \cite{sarovar2020detecting,rudinger2021crosstalk}.

We first describe the diagnostic checks and encodings, then construct their structural references and apply them to IBM Kingston data. Auxiliary width certificates, a 600-instance structural ensemble and a matched-sector mixer control are reported in the Supplemental Material \cite{SMV7}.

\section{Diagnostic procedure and benchmark encoding}
\label{sec:axis}

The procedure uses three checks, applied to the circuit and sampling budget under study.
\begin{enumerate}
\item \emph{Validate the encoded target.} Compare encoded minima with certified feasible optima. If they differ, report surrogate energy and target-set probability as distinct observables.
\item \emph{Calculate an ideal reference.} Identify support-graph blocks or another justified classical representation and evaluate the executed ansatz point. For an event of ideal probability $P_{\rm id}$, compare $N_sP_{\rm id}$ with one before interpreting its absence in $N_s$ shots.
\item \emph{Resolve what aggregates conceal.} Match cardinality sectors where appropriate, compare energy with a product of measured bit marginals, and decompose weighted covariance by logical block. Propagate the joint counting uncertainty when comparing a block with the aggregate containing it.
\end{enumerate}
These checks concern the recorded execution. A conditional ideal reference remains valid for fixed angles even when their optimization history is incomplete; claims about optimizer performance require that history separately.

The Legal Amazon instance uses the 773 municipality polygons in the official IBGE 2024 product \cite{IBGE2024AmazoniaLegal}.  Geometric centroids are computed after projection to SIRGAS 2000/Brazil Polyconic (EPSG:5880) and transformed to longitude--latitude; PEF base coordinates come from the Army registry \cite{MinistDefesa2023}.  The modeled distance is the haversine great-circle distance
\begin{equation}
  \label{eq:haversine}
  d(i,j) = 2r_\oplus \arcsin\!\sqrt{\sin^2\!\tfrac{\theta_j-\theta_i}{2} + \cos\theta_i\cos\theta_j\sin^2\!\tfrac{\phi_j-\phi_i}{2}},
\end{equation}
where $\theta_i$ and $\phi_i$ are, respectively, the latitude and longitude of site $i$ in radians, and $r_\oplus=6371$ km.  The coverage graph $G=(V,E_R)$ has edge $(i,j)$ if and only if $d(i,j)\le R$; the mean degree grows as $\bar d\propto R^{1.73}$ over the municipality set (a log--log fit across nine radii, residual standard deviation $0.07$), not as the $R^2$ of a uniform planar process.  The municipality centroids are spatially nonuniform; the fit is a descriptive property of this benchmark.

Three frozen benchmarks separate encoding size from geographic scale.  PEF-24 uses 24 frontier bases and 23 synthetic waypoints at $R=200$ km; its optimum has ten bases.  Arco-79 uses 79 published municipality centroids and 49 reachable synthetic targets at $R=100$ km; its optimum has 31 bases.  Arco-79 is a frozen resource benchmark; its coordinate list is reproducible, while its unrecorded original selection rule precludes treating it as a statistical sample.  Legal-Amazon-773 uses all official municipality centroids as both candidates and targets at $R=200$ km; the constrained dominating-set optimum is 47. Its associated bare-register quadratic surrogate is not a faithful encoding, and 47 is not asserted to be its ground-state optimum.  Only the first two were executed.  Construction details are given in the Supplemental Material \cite{SMV7}; the frozen inputs and hardware archives are included in the versioned code-and-data release (see Data and code availability). Table~\ref*{sm:tab:sm_families} distinguishes these benchmarks and the auxiliary families.

With distinct base set $B$ ($|B|=\nB$) and task set $T$ ($|T|=\nt$), define the coverage matrix $A\in\{0,1\}^{\nt\times\nB}$, $A_{tb}=1$ if and only if $d(b,t)\le R$.  The minimum set cover is $\min_x \mathbf 1^\top x$ subject to $Ax\ge\mathbf 1$, and the compact quadratic unconstrained binary optimization (QUBO) surrogate studied here is
\begin{equation}
  \label{eq:qubo_msc}
  H_C^{\mathrm{sur}}=\sum_b x_b+\lambda\sum_t\Big(1-\sum_b A_{tb}x_b\Big)^2.
\end{equation}
Because $x_b^2=x_b$ and $A_{tb}^2=A_{tb}$, expanding the square and collecting the diagonal into the linear part gives the QUBO coefficients
\begin{equation}
  \label{eq:qubo_coeffs}
  q_b=1-\lambda\sum_t A_{tb},\qquad
  Q_{bb'}=2\lambda\sum_t A_{tb}A_{tb'}\ \ (b<b'),
\end{equation}
plus the constant $\lambda\nt$.  The Ising form follows from $x_b=(1-Z_b)/2$, giving $J_{bb'}=Q_{bb'}/4$ and $h_b=-q_b/2-\tfrac14\sum_{b'\neq b}Q_{bb'}$.

Equation~(\ref{eq:qubo_msc}) is a \emph{two-sided surrogate} and we use it as such.  The squared penalty penalizes deviations from exact-one coverage and is therefore not a faithful encoding of the inequality $Ax\ge\mathbf 1$.  Its ground state need not be a minimum cover: on the frozen 24-base instance, exact enumeration puts 144 infeasible weight-nine states below the 288 optimal covers [Sec.~\ref{ssec:kingston1}].  We adopt it for the bare-register 24- and 79-qubit comparisons because the faithful encoding cannot fit the same 24-qubit register; we then execute its 51-qubit slack construction separately.

The faithful construction uses private slack bits,
\begin{equation}
 H_C^{\rm faith}=\sum_bx_b+\lambda\sum_t\big[(Ax)_t-1-s_t\big]^2,
 \label{eq:faithful}
\end{equation}
with $s_t=\sum_{j=0}^{\ell_t-1}2^j y_{tj}$ and $\ell_t=\lceil\log_2(\deg t)\rceil$. Values $0,\ldots,\deg t-1$ represent every feasible residual; extra binary values cannot satisfy an equality. For feasible coverage instances with unit base costs, private slack and no additional constraints, $\lambda>1$ is sufficient for ground-state faithfulness. An infeasible base assignment with $u$ uncovered targets can be repaired with at most $u$ extra bases, while its penalty is at least $\lambda u$; resetting slack then lowers energy. The same repair argument handles strict multiplicities using summed deficits [Sec.~\ref*{sm:sm:structure}]. A one-target, one-base instance can tie at $\lambda=1$. The executed value $\lambda=10$ was fixed in the historical circuits; exact block enumeration independently checks the constructed operators.

Private binary slack supplies a quadratic, ground-state-faithful encoding of the same coverage task, with coefficients and block probabilities that can be checked exactly. It is a reference construction, not a claim of minimal circuit resources. Direct inequality implementations with qudits \cite{bottarelli2025inequality} and oracle-based indicator-function QAOA \cite{bucher2025penaltyfree} offer alternative encodings with different registers and gate requirements. Their performance comparisons do not establish which construction is preferable for this executed qubit circuit; that would require matching the objective, compiled resources and training protocol.

The slack bits of a target couple only to the bases covering that target, and those bases already lie in one component of $G_{ZZ}$, so the ancillas join a block instead of bridging two: \emph{the block count is unchanged}.  What grows is the register
and the coupling count.  On the 24-base instance the faithful operator needs 51 qubits against 24 and $\nzz=103$ against $17$, still splitting into 13 blocks, now of at most 13 variables, with combined block-state dimension $8958$ against $2^{51}$; on the 79-base instance it needs 89 qubits against 79 and $\nzz=29$ against $9$, still 70 blocks, with combined dimension 220.  The faithful encoding is larger but preserves the block partition; routing on the device is examined in Sec.~\ref{ssec:kingston1}.

Mapping the QUBO to an Ising cost Hamiltonian $H_C=\mathrm{const}+\sum_b\tilde h_b Z_b +\sum_{b<b'}\tilde J_{bb'}Z_bZ_{b'}$ gives one logical $ZZ$ interaction for each nonzero $\tilde J_{bb'}$. The interaction count is
\begin{equation}
  \label{eq:nzz}
  \nzz = \big|\{(b,b'):b<b',\,\tilde J_{bb'}\neq0\}\big|.
\end{equation}
The sign of a coupling does not change the gate it requires: the bare-register surrogate has only positive couplings, whereas 76 of the 103 PEF-24 faithful-encoding couplings are negative base--slack terms.  The standard two-entangling-gate synthesis of each $ZZ$ rotation gives a pre-routing count $2p\nzz$ for these circuits; compilation and routing determine the executed count. A gate-error product is only a resource proxy, not a measured fidelity.  For the bare-register quadratic surrogate, let $C_t=\{b\in B:A_{tb}=1\}$ and $m_t=|C_t|$.  Its interaction set is $E_{ZZ}^{\rm bare}=\bigcup_{t\in T}\binom{C_t}{2}$, and hence
\begin{equation}
  \label{eq:sparsification}
  n_{ZZ}^{\rm bare}
  =\left|\bigcup_{t\in T}\binom{C_t}{2}\right|
  \le \sum_{t\in T}\binom{m_t}{2}.
\end{equation}
Repeated co-covering pairs are counted once on the left and may occur several times on the right.  Equation~(\ref{eq:sparsification}) does not count slack couplings.  For a threshold $t_t$, an equality with $\ell_t=\lceil\log_2(m_t-t_t+1)\rceil$ private slack bits makes $C_t\cup S_t$ a clique, giving the separate upper bound $n_{ZZ}^{\rm slack}\le\sum_t\binom{m_t+\ell_t}{2}$ before repeated base--base pairs are removed.  For PEF-24 the bare clique sum and distinct count are $40$ and $17$, while the slack clique sum and distinct count are $126$ and $103$. For Arco-79 the corresponding four values are $10$, $9$, $30$ and $29$.  The dense 773-base bare-register dominating-set surrogate has $\nzz=39{,}974$ logical interactions before routing.

Gate count alone is insufficient even at fixed register size, since placement and scheduling change the error exposure. We use a first-order resource proxy with independent gate and readout factors,
\begin{equation}
 \label{eq:resource}
 \Fest\approx(1-\epsilon_{2Q})^{n_{2Q}}(1-\epsilon_{\mathrm{ro}})^{\nq},
\end{equation}
where $\nq$ counts measured qubits, including slack ($\nq=51$ but $\nB=24$ for faithful PEF-24). This extends the gate-count proxy used by Sharma and Lau \cite{sharma2026circuits} with a measured-qubit readout factor. Both products assume independent contributions; they omit idling and correlated errors and are not measured process fidelities. For the executed circuits rather than device-wide averages,
\begin{equation}
  \label{eq:resource_exact}
  \Fest=\prod_{e\in\mathcal E}\big(1-\epsilon_{2Q}^{(e)}\big)^{m_e}\prod_{q\in\mathcal Q}\big(1-\epsilon_{\mathrm{ro}}^{(q)}\big),
\end{equation}
over the couplers $\mathcal E$ of the transpiled circuit, each counted with the number $m_e$ of two-qubit gates it carries, and the qubits $\mathcal Q$ that are measured. Section~\ref{sec:rung2} shows that the readout factor reverses the ordering the gate count alone predicts.

\section{Support graph and classical simulability}
\label{sec:factorise}

\subsection{Exact block factorization}
\label{ssec:blocks}
The factorization is algebraic. Geometry can provide sufficient conditions for it, but is not required.  Write a diagonal cost as $H_C=\sum_\alpha H_\alpha$ and define its \emph{support graph} $G_{\rm sup}$ on the variables by joining two variables whenever they both occur in the support of some nonzero term $H_\alpha$.  Let $V_1,\ldots,V_M$ be its connected components.

\textbf{Block-locality proposition.}
Suppose every cost and mixer layer is a tensor product of unitaries supported within the same sets $V_i$.  Then, for every depth $p$ and independently of the input state, the layered unitary satisfies
\begin{subequations}
\begin{equation}
  \label{eq:unitary_factor}
  U_p(\boldsymbol\gamma,\boldsymbol\beta)
  =\bigotimes_i U_p^{(i)}(\boldsymbol\gamma,\boldsymbol\beta).
\end{equation}
If, in addition, the initial state is a product across that partition,
$\ket{\psi_0}=\bigotimes_i\ket{\psi_0^{(i)}}$, then
\begin{equation}
  \label{eq:factor}
  \ket{\psi_p(\boldsymbol\gamma,\boldsymbol\beta)}=
  \bigotimes_i\ket{\psi_p^{(i)}(\boldsymbol\gamma,\boldsymbol\beta)}.
\end{equation}
\end{subequations}
Each block state $\ket{\psi_0^{(i)}}$ may itself be entangled among the qubits in $V_i$; the requirement is separability between blocks, not between individual qubits.  In particular, block-local layers cannot generate entanglement between blocks that are initially unentangled.

\emph{Proof.} Each cost term lies in one component, so
\begin{equation}
 \label{eq:kron}
 H_C=\sum_i H_C^{(i)}\otimes\mathbb 1_{\bar V_i},
 \quad e^{-i\gamma H_C}=\bigotimes_i e^{-i\gamma H_C^{(i)}}.
\end{equation}
Mixer layers factor by hypothesis. Multiplication preserves the common tensor-product partition, proving Eq.~(\ref{eq:unitary_factor}); applying it to the product input proves Eq.~(\ref{eq:factor}). $\square$

If the initial state is entangled across the partition, Eq.~(\ref{eq:unitary_factor}) still holds but Eq.~(\ref{eq:factor}) does not: block-local unitaries preserve the Schmidt coefficients across every block cut and therefore cannot remove the pre-existing cross-block entanglement.  The joint output distribution then need not be the product of its block marginals.  A low cross-block Schmidt rank may nevertheless leave a weaker tensor-network simulation efficient, as the Dicke-state example below illustrates.

For the set-cover QUBO, every target contributes a clique on its cover set, so every coverage term lies within one component of the interaction graph $G_{ZZ}$, which for this QUBO coincides with $G_{\rm sup}$.  The unconstrained objective is additive across the components and its optimum is obtained blockwise.  The combined block-state dimension $D_{\rm block}=\sum_i2^{|V_i|}$ describes this representation, not a runtime or a lower bound on simulation cost.  The same statement holds for the slack encoding because each target's slack bits attach only to its existing cover set's block.

Geometry gives a useful corollary.  In any metric radius-$R$ coverage instance, two bases that co-cover a target satisfy
\begin{equation}
  \label{eq:2r_corollary}
  \begin{gathered}
  d(b,b')\le d(b,t)+d(t,b')\le2R,\\
  E(G_{ZZ})\subseteq E(G_B(2R)).
  \end{gathered}
\end{equation}
Thus groups separated by more than $2R$ cannot share a cost term and may form distinct blocks.  The graph $G_B(2R)$ is a distance-threshold proximity graph; $G_{ZZ}$ is only a subgraph of it, because a nearby pair may fail to share a target.  Treewidth is monotone under taking subgraphs, so every upper bound established for $G_B(2R)$ also bounds $G_{ZZ}$.  Geometry is not necessary for factorization, and geometric locality alone may produce a connected path with low width but no exact blocks.

The edge count is a weak diagnostic: a path on 24 variables has 23 interactions and one component, whereas four disconnected six-variable cliques have 60 and factorize exactly.  The scope audit reproduces the block identity on 250 random incidence hypergraphs and verifies Eq.~(\ref{eq:2r_corollary}) on 768 spatial instances; these computational checks guard the implementation, while the proof above supplies the general result.

The 24-qubit circuit run on hardware in Sec.~\ref{sec:rung1} has $\nzz=17$ and splits into 13 components whose largest block has four bases: the combined block-state dimension is 62, against $2^{24}=1.7\times10^{7}$ for the register it occupies.  The 79-qubit instance splits into 70 components of at most three bases, with combined dimension 160.

Standard QAOA satisfies both forms of the theorem because $\ket{+}^{\otimes\nB}$ is a product state and the transverse-field mixer is a product of single-qubit rotations.  A circuit with blocks $\{V_i\}$ is therefore $M$ independent circuits on at most $\max_i|V_i|$ qubits.  For PEF-24, the ideal 24-qubit state agrees with the tensor product of 13 block states to double-precision round-off, and the 288 optimal covers are exactly the Cartesian product $2\times4\times3\times3\times2\times2$ of the per-block optima.  A mixer gate crossing the partition would merge the corresponding circuit blocks.  Section~\ref{ssec:kingston1} draws the consequence for the
hardware run, and Supplemental Sec.~\ref*{sm:sm:width_certificates} examines what replaces Eq.~(\ref{eq:unitary_factor}) when the
operator stops decomposing.

The product-state conclusion does not apply directly to the Dicke input used by the cardinality-preserving control.  The operator still factorizes because its $XY$ mixer uses the same within-block edge support, while the Dicke state has cross-block Schmidt rank at most $k+1$.  The Schmidt spectrum, conditional-sector simulation and block-weight ceiling on $P(\mathrm{opt})$ are derived in Sec.~\ref*{sm:sm:dicke}; the low rank preserves efficient simulation of the frozen PEF-24 control.

\subsection{Scope beyond disconnected circuits}
The exact reference used below follows from disconnected support and product input. When components merge, a verified width bound may still permit contraction at fixed depth, but component size and logical gate count alone do not certify simulation difficulty. Supplemental Sec.~\ref*{sm:sm:width_certificates} gives the strip construction, depth-dependent contraction bounds and their geometric assumptions. The 773-base surrogate has $112\le\mathrm{tw}\le151$ and supplies no small-block reference of the type used for the hardware experiments.

A separate 600-instance ensemble varies register size and radius on official municipality sites. Larger support blocks are associated with more interactions, including within fixed size--radius cells, but the association is specific to the generator and does not constitute a runtime law. The ensemble, fixed-cell uncertainty and figure are in Supplemental Sec.~\ref*{sm:sm:ssec:ensemble}.

\section{Exact-reference hardware diagnostics}
\label{sec:rung1}

\subsection{PEF-24 and its exact ideal reference}
\label{ssec:pef}
\begin{figure*}[t]
  \centering
  \includegraphics[width=\linewidth]{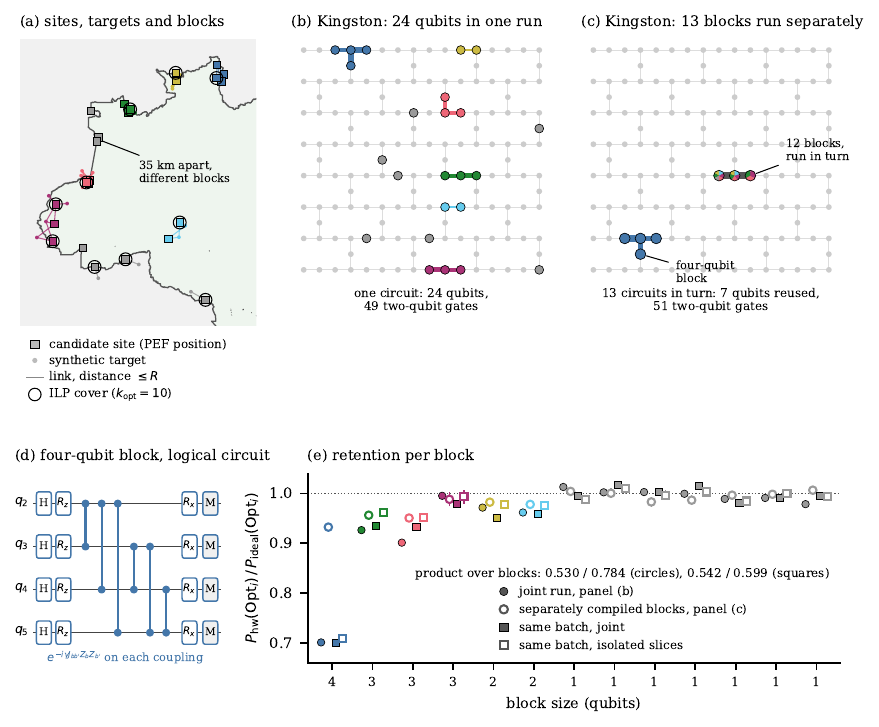}
  \caption{PEF-24 from geography to hardware.  (a) Bases, synthetic waypoints, one certified minimum cover and the 13 support-graph blocks.  (b) The joint $p=1$ Kingston layout; all 49 physical two-qubit gates remain within the logical blocks.  (c) Separately transpiled block circuits.  (d) The four-qubit logical block.  (e) Measured-to-ideal optimal-block retention: filled/open circles are joint/separately transpiled runs ($10^5$/$10^4$ shots); filled/open squares are same-job joint/slice runs ($10^5$/$10^4$ shots). Bars are binomial standard errors conditional on aggregate counts.  Complete job, placement and uncertainty details are given in the Supplemental Material.}
  \label{fig:blocks}
\end{figure*}

The 24 PEF bases span 2,005 km N--S along the frontier [Fig.~\ref{fig:blocks}(a)].  At $R=200$ km the ILP selects $\kopt=10$ active bases covering all 23 synthetic waypoints.  Eleven waypoints fall outside the IBGE boundary, up to 92 km across the border; the draw is a geometric benchmark, not an operational siting.  It was accepted only if feasible and if the greedy cover size equaled ten, so $k_\mathrm{greedy}=\kopt$ is a selection criterion rather than evidence about greedy in general.  All results refer to this published frozen instance [Sec.~\ref*{sm:sm:details}].

\subsection{Measured concentration and loss decomposition}
\label{ssec:kingston1}
On IBM Kingston (Heron r2, 156 qubits), the PEF-24 $p=1$ QAOA circuit ($\nzz=17$; $\gamma^*=-0.04$, $\beta^*=0.35$) uses 34 logical two-qubit gates and transpiles to depth 64 and 49 physical two-qubit gates.  Over $10^5$ shots with XY4 dynamical decoupling and Pauli twirling, it samples a minimum cover ($k=10$, full coverage) 79 times: $\Phw=7.90\times10^{-4}$, with Wilson 95\% interval $[6.34,9.84]\times10^{-4}$.  The optimum is $288$-fold degenerate.  Because the QAOA state explores the full 24-bit register, the null is $288/2^{24}=1.7166\times10^{-5}$, and the all-register concentration is
\begin{equation}
  \eta_{\rm all}=\frac{P(\mathrm{opt})}{288/2^{24}}=46.02,
  \qquad 95\%\ \mathrm{CI}=[36.93,57.34].
\end{equation}
As a separate cardinality diagnostic, 15,351 shots have $k=10$; conditioning both numerator and null on that sector gives $\eta_{k=10}=35.05$ with Wilson interval $[28.14,43.64]$.  An unconditional probability must not be divided by this conditional null; the two estimands are reported separately.  The mean QUBO energy falls from the exact uniform-bitstring value $E_\mathrm{rnd}=187.0$ to $127.4$, a $31.9\%$ reduction.

Interpreting that number as evidence of optimization performance requires applying Eqs.~(\ref{eq:unitary_factor}) and (\ref{eq:factor}).  This instance has 13 blocks of at most four bases, the mixer is a product of $R_x$ rotations and the initial state is $\ket{+}^{\otimes24}$, so the ideal logical unitary submitted on 24 qubits is equivalent to a tensor product of 13 block circuits and its output state is a product across them.  It cannot create ideal correlations between the blocks, although this algebraic statement does not exclude correlations from hardware noise or crosstalk.  The state vector confirms this, and the optimum factorizes with the state (Sec.~\ref{ssec:blocks}).

The decomposition limits the benchmark but also supplies an exact noiseless reference.  The block calculation gives $P_{\rm ideal}(\mathrm{opt})=1.8254\times10^{-3}$, or $\eta_{\rm all}^{\rm ideal}=106.34$; the hardware-to-ideal probability ratio is $0.4328$.  Every minimum cover has $k=10$, so this ratio factorizes exactly into a loss of cardinality-sector mass and a loss of selection within the sector,
\begin{equation}
  \label{eq:sector_split}
  \frac{\Phw(\mathrm{opt})}{P_{\rm ideal}(\mathrm{opt})}
  =\underbrace{\frac{0.1535}{0.1747}}_{P(k=10)}\times
   \underbrace{\frac{5.15\times10^{-3}}{1.045\times10^{-2}}}_{P(\mathrm{opt}\mid k=10)}
  =0.879\times0.492 .
\end{equation}
The ideal in-sector concentration is $\eta_{k=10}^{\rm ideal}=71.2$ against the measured $35.05$.  The device retained $88\%$ of the weight-ten sector but only $49\%$ of the ideal selection probability within it, so $85\%$ of the logarithmic deficit is intra-sector.  Because the ideal state is a product over 13 blocks, multiplying measured per-block optimal probabilities supplies a block-product null.  It predicts 96.7 hits against 79 observed; an independent-block permutation test gives $p=0.07$, and no block-indicator correlation exceeds $0.01$ in magnitude.  The four-qubit block supplies the largest observed within-block contribution, but block identity, two-qubit-gate count, placement and schedule are confounded.  The residual cross-block retention is $0.82$ with bootstrap interval $[0.65,1.00]$, compatible with no cross-block deficit and with loss as large as $35\%$; no such deficit is resolved.  Its numerical equality to $\Fest=0.817$ is coincidental.

As established in Sec.~\ref{sec:axis}, the target covers are not ground states of the implemented surrogate.  The counts put 72 shots on its 144 infeasible ground strings, a concentration of $83.9\times$ over uniform against $46.0\times$ for the target covers.  This exposes the encoding mismatch; it does not show that the fixed pilot angles optimized the surrogate or that the hardware found its ground state.

At $p=2,3$, tiled rather than re-optimized angles reduce both ideal and measured target concentration.  The exact reference therefore identifies this depth trend as an ansatz-point effect rather than evidence of decoherence; full values are in the Supplemental Material.

\subsubsection{Observables and controlled contrasts}
For job $e$, round $r$ and block $i$, let $p^J_{eri}$ and $p^S_{eri}$ be optimal-block probabilities in the joint circuit and its isolated same-placement slice. Over the six blocks $\mathcal G$ containing two-qubit gates, define
\begin{equation}
 R_{er}=\prod_{i\in\mathcal G}\frac{p^J_{eri}}{p^S_{eri}},\qquad
 R_e=\exp\!\left[\tfrac14\sum_{r=1}^4\log R_{er}\right].
 \label{eq:joint_diagnostic}
\end{equation}
For the faithful circuit let $E$ be its mean QUBO energy and $E_{\rm prod}$ the energy of independent bits with the same marginals. The covariance diagnostic is
\begin{equation}
 C_{er}=\frac{E_{{\rm prod},er}-E_{er}}{E_{\rm prod}^{\rm id}-E^{\rm id}},
 \qquad C_e=\tfrac14\sum_{r=1}^4 C_{er}.
 \label{eq:correlation_diagnostic}
\end{equation}
For $E(x)=c+\sum_iq_ix_i+\sum_{i<j}Q_{ij}x_ix_j$,
\begin{equation}
 E_{\rm prod}-E=-\sum_{i<j}Q_{ij}\operatorname{Cov}(x_i,x_j).
 \label{eq:covariance_diagnostic}
\end{equation}
Thus $C$ compares a weighted covariance moment with its exact ideal value. It can involve cancellations and lie outside $[0,1]$; it is neither state fidelity nor an entanglement witness. Normalization requires a nonzero ideal denominator, and becomes sensitive to perturbations when that denominator is small.

Exact synthetic output controls clarify its use [Sec.~\ref*{sm:sm:diagnostic_controls}]. Removing all covariances from the ideal faithful distribution leaves a $\ControlProductReduction\%$ uniform-reference energy reduction, against $48.6\%$ ideally. Under independent asymmetric output flips with $p_{01}+p_{10}=0.2$, varying the bias changes energy reduction from $\ControlBiasReductionMin\%$ to $\ControlBiasReductionMax\%$ while $C=0.64$ throughout. Energy can therefore appear improved despite covariance attenuation. Conversely, a four-bit example preserves every block marginal and $C=1$ while doubling the between-block optimum factor. The latter requires a separate dependence diagnostic, such as the measured global optimum probability divided by its block-marginal product. These are exact controls of output distributions, not identified hardware error mechanisms. Table~\ref{tab:decisions} summarizes the complementary decisions.

\begin{table}[t]
\caption{Complementary decisions supplied by the reference audit. Each comparison has a distinct target and scope.}
\label{tab:decisions}
\begin{ruledtabular}
\begin{tabular}{p{0.28\columnwidth}p{0.64\columnwidth}}
Comparison & Decision supported \\
\hline
Encoded minima vs. constraints & Whether ground-state success represents the intended task \\
Ideal optimum mass vs. shots & Whether zero optimum counts can diagnose measured loss \\
Energy and $C$ vs. exact references & Whether energy change accompanies weighted covariance, and which block moments the aggregate masks \\
Global vs. block-product optimum mass & Whether optimal-block events depart from the product null \\
Joint vs. isolated slices, $R$ & How execution context changes local success; mechanism remains confounded \\
\end{tabular}
\end{ruledtabular}
\end{table}

\subsubsection{Standardized fixed-circuit campaign}
\begin{figure*}[t]
  \centering
  \includegraphics[width=\linewidth]{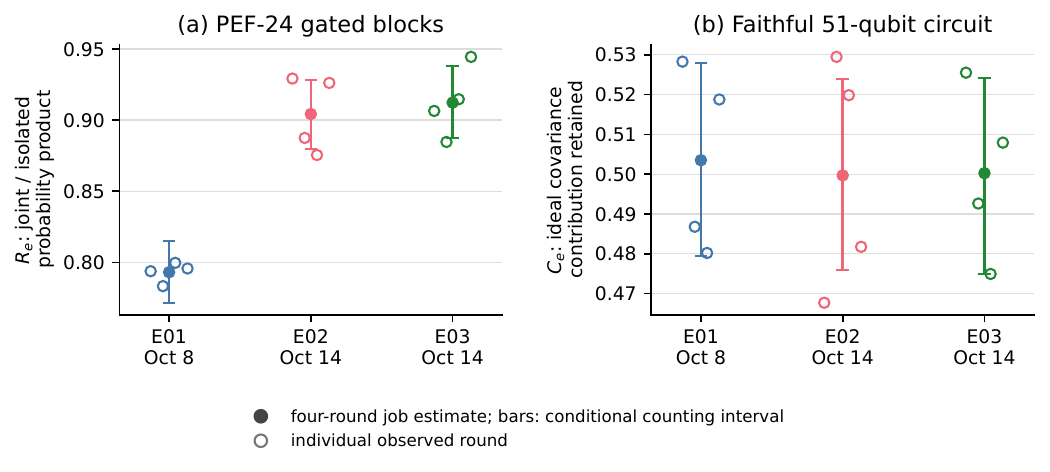}
  \caption{Two endpoints in three fixed-circuit jobs: (a) $R_e$ from Eq.~(\ref{eq:joint_diagnostic}); (b) $C_e$ from Eq.~(\ref{eq:correlation_diagnostic}). Filled points are four-round job estimates; open points are the four observed rounds. Bars are conditional counting-model 95\% percentile intervals from 4000 bootstrap draws, holding those rounds fixed. They do not include drift, unarchived twirling-stratum variability or between-job uncertainty. Variance subtraction is reported separately, not folded into these bars.}

  \label{fig:threecal}
\end{figure*}

Three jobs reused one ISA-circuit bundle without re-transpilation. Each contained four randomized rounds of the joint circuit, its 13 same-placement slices and the matched-mitigation faithful circuit. Figure~\ref{fig:threecal} gives $R_e=0.7931,0.9043,0.9123$ and $C_e=0.5035,0.4997,0.5002$, with counting-model intervals conditional on the four observed rounds. The positive descriptive between-job excess of log $R$ is driven by E01: omitting that job gives observed dispersion only $0.19$ times the remaining mean working estimation variance and zero truncated excess. No excess is resolved for $C$. These endpoint-specific descriptions use different circuits and do not compare intrinsic stability. Full variance components, job-omission checks and a hypothetical variance-inflation sensitivity are in Sec.~\ref*{sm:sm:threecal}.

Earlier block-wise jobs provide secondary sensitivity checks rather than coequal evidence.  In one interleaved job, the geometric means of four round-level products were $0.877$ and $0.671$ for two compilations.  A separately transpiled placement of the four-qubit block retained $0.93$ against $0.69$ and $0.65$ in the joint compilations, but placement, routing, schedule and compilation are confounded.  Full per-block tables and historical jobs are retained in the Supplemental Material.

\subsubsection{Faithful encoding and routing}
\textbf{Faithful-encoding execution.} We also executed the faithful slack encoding on IBM Kingston. As shown in Sec.~\ref{sec:axis}, this formulation requires 51 qubits and $\nzz=103$ interactions, yet still decomposes into 13 blocks. It therefore fits within Kingston's 156-qubit register. The executed angles were selected by minimizing $\langle H\rangle$, not by maximizing optimum probability.  At those angles, the total ideal probability of the set of complete optimum strings is $5.47\times10^{-20}$, so zero optimum counts in $10^5$ shots are expected even ideally and do not diagnose the device.  Neither execution produced a full 51-bit string satisfying every slack equality (Wilson upper 95\% bound $3.84\times10^{-5}$); the observed base-register cover fractions, $0.53569$ and $0.4326$, are marginal diagnostics rather than feasibility probabilities of the encoded problem.  Expected-energy reduction is likewise dominated by single-bit marginals: a product distribution with the measured marginals reproduces about $90\%$ of the measured reduction.  A stronger classical null preserving the base-cardinality histogram and slack marginals has energy $239.45$, below the measured $245.56$ in the first run. Accordingly, positive $E_{\rm prod}-E$ diagnoses preservation of the ideal weighted covariance, rather than optimization superiority over structured classical distributions; the two runs retain $56\%$ and $50\%$ of its ideal contribution.  The executions used different transpilations, placements, dates and calibrations, so they are replications under different runtime settings rather than a causal estimate of dynamical decoupling or twirling.  Their enhancement relative to cardinality-matched nulls is similar ($1.198$ and $1.155$), indicating that the raw cover-rate change is largely associated with the shift in measured base cardinality.

\begin{figure*}[t]
  \centering
  \includegraphics[width=\linewidth]{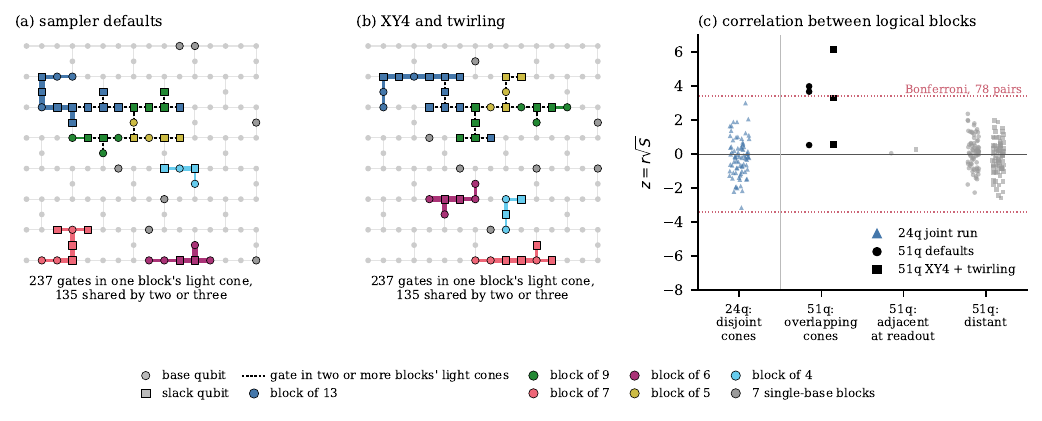}
  \caption{The faithful 51-qubit circuit on IBM Kingston.  (a, b) Executed layouts of the two runs (two separate transpilations), qubits colored by logical block at measurement (circles: base qubits; squares: slack qubits).  A two-qubit gate is drawn in block color when it lies in the backward light cone of that block only, and dashed when it lies in the cones of two or more blocks; the cones follow every routing SWAP.  (c) Correlation $z=r\sqrt{N_s}$, with $N_s$ the number of shots, between the per-shot energies of every pair of logical blocks: for the 24-qubit joint run, where all cones are disjoint, and for the 51-qubit runs, grouped into pairs with overlapping cones, pairs with disjoint cones whose measured qubits are adjacent on the chip, and the rest.  Dotted lines: the Bonferroni threshold for 78 pairs.  Data: \texttt{ibm\_circuits/exact51\_*.ops.json}, \texttt{exact51\_cross\_block.json}.}
  \label{fig:exact51}
\end{figure*}

The faithful circuit also tests block locality after routing.  Gate-mediated propagation between two logical blocks requires overlap of their backward light cones, although correlated readout and other nonlocal noise need not.  In each 51-qubit transpilation three of the 78 block pairs have overlapping cones.  Two of those pairs carry the largest observed correlations, while no disjoint-cone pair passes the stated Bonferroni threshold; the third overlapping pair is not correlated.  This exploratory association recurs across only two transpilations, is weak ($r\le0.02$) and is not a deterministic consequence of routing [Fig.~\ref{fig:exact51}].

The historical executions are summarized in Table~\ref*{sm:tab:kingston}; the standardized campaign is the primary repeated-job comparison.

\subsubsection{What the global covariance ratio conceals}
\label{ssec:block_profiles}
For a logical component $V_\alpha$, restrict Eq.~(\ref{eq:covariance_diagnostic}) to its internal pairs and call the result $\Delta_\alpha$. The exact ideal reference gives
\begin{equation}
 \begin{aligned}
 C_\alpha&=\Delta_\alpha/\Delta_\alpha^{\rm id},\\
 C_{\rm global}&=\sum_\alpha w_\alpha C_\alpha,\\
 w_\alpha=\Delta_\alpha^{\rm id}/\sum_\nu\Delta_\nu^{\rm id}.
 \end{aligned}
 \label{eq:block_profile}
\end{equation}
The six interacting components have nonzero ideal denominators; singletons contribute zero and are not normalized. All weights are positive here. Components with 9 and 13 variables carry $\BlockLargeWeight\%$ of the global denominator. A five-variable component carries only $\BlockFiveWeight\%$, so its local discrepancy can contribute little to the aggregate.

\begin{figure*}[t]
 \centering
 \includegraphics[width=\linewidth]{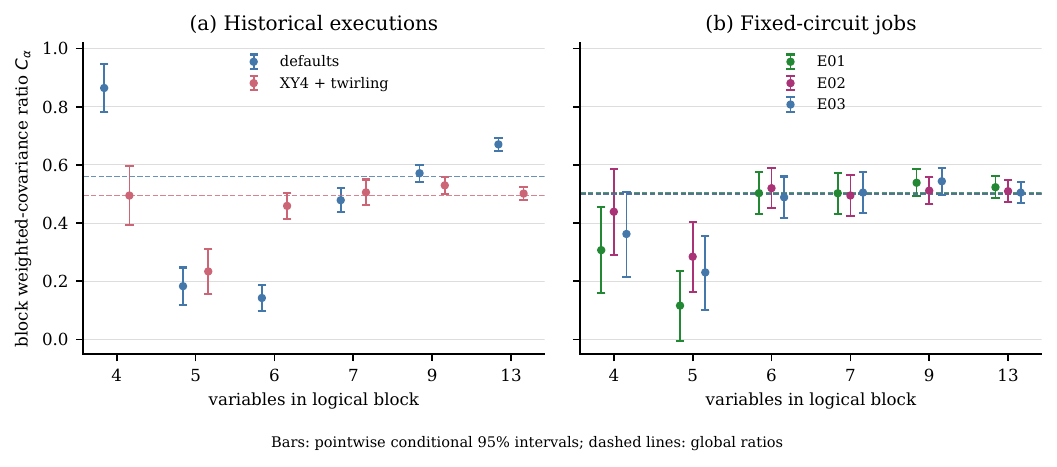}
 \caption{Weighted-covariance profiles of the faithful circuit: (a) historical executions; (b) three fixed-circuit jobs. Points are $C_\alpha$ and bars are pointwise conditional 95\% percentile intervals from 10,000 joint full-bitstring bootstrap draws per pub; job statistics average the four fixed rounds. Dashed lines show the respective global ratios. The five-variable block carries only $\BlockFiveWeight\%$ of the ideal global contribution, yet its block-minus-global contrast is below zero in every context under simultaneous bands for all 30 contrasts (Supplemental Material). Pointwise bar overlap is not the contrast test. Different historical transpilations and settings do not isolate mitigation effects.}
 \label{fig:block_profiles}
\end{figure*}

Figure~\ref{fig:block_profiles} resolves that discrepancy. Its five-variable ratios are $\BlockFiveDefaults$ and $\BlockFiveMatched$ in the historical runs and $\BlockFiveEOne$, $\BlockFiveETwo$, $\BlockFiveEThree$ in E01--E03, against global ratios near $0.50$ in the campaign. Define $G_\alpha=C_\alpha-C_{\rm global}$. Approximate simultaneous 95\% bands over all 30 block/context contrasts put the five-variable $G_\alpha$ below zero in all five contexts. In E01--E03 the bands are $[\BlockFiveEOneLow,\BlockFiveEOneHigh]$, $[\BlockFiveETwoLow,\BlockFiveETwoHigh]$ and $[\BlockFiveEThreeLow,\BlockFiveEThreeHigh]$. An aggregate ratio near one half therefore does not describe uniform local attenuation. This is a discrepancy in a weighted output moment, not a determination of individual covariances or of an error process.

The exploration was motivated by historical point estimates. Its 10,000 full-bitstring multinomial draws per pub preserve joint estimation covariance between blocks and condition on the observed rounds. The new campaign intervals are Monte Carlo realizations of the same counting model as Fig.~\ref{fig:threecal}; their central estimands are unchanged. They do not recover twirling-stratum variability or drift, and differences between historical placements and mitigation settings remain confounded. Methods, round estimates and the full contrast family are in Sec.~\ref*{sm:sm:block_profiles}. This calculation uses 51 bit means and six quadratic block statistics, in addition to the exact block reference, and requires no full 51-qubit state vector or new hardware shots.

A provenance audit maps all six interacting blocks to their measured physical qubits and backward gate cones, then joins these records to the retrospectively recovered E01--E03 job properties. For the five-variable block, the gate-count-weighted mean CZ errors are $0.226\%$, $0.229\%$ and $0.229\%$, while mean measured-qubit readout errors are $1.538\%$, $0.947\%$ and $0.771\%$. Qubit 30 has a recorded readout error of $4.663\%$ in E01, falling to $0.854\%$ and $0.732\%$. However, the block's point covariance ratio is highest in E02, and the recorded CZ errors on all used couplers are unchanged between E02 and E03. Applying independent asymmetric readout at the archived per-qubit rates to the exact five-variable distribution gives $C_\alpha^{\rm ro}=\ReadoutCOne,\ReadoutCTwo,\ReadoutCThree$. Its predicted attenuation is only $\ReadoutFractionMin$--$\ReadoutFractionMax\%$ of the observed deficit $1-C_\alpha$. These are model-conditional point comparisons, not causal fractions or confidence bounds: independent readout at the recorded rates alone cannot reproduce the local deficit. The remaining mechanism is not identified. Supplemental Sec.~\ref*{sm:sm:physical_audit} reports all six blocks, parameter provenance and the limits of this descriptive three-job comparison.

\subsection{Arco-79 as an ansatz null}
\label{sec:rung2}
Arco-79 has 49 reachable synthetic targets at $R=100$ km and optimum $\kopt=31$.  Thirty-nine targets have degree one and ten have degree two, giving only $\nzz=9$ logical interactions despite the 79-qubit register.  The instance decomposes into 70 blocks of at most three bases, so its ideal distribution is computed exactly by block convolution.  An oracle-assisted $p=1$ search over the tested angular domain raises $P_{\rm ideal}(\mathrm{opt})$ from $9.9\times10^{-20}$ at transferred angles to a largest found value of $2.0\times10^{-9}$.  The own-angle run lowers mean cardinality from $42.8$ to $31.9$ but produces no minimum-cover hit. Even the largest found ideal probability requires about $5\times10^8$ shots per expected optimum, so zero hits in the $10^5$-shot own-angle run do not diagnose the device and are not converted into a concentration ratio.

The circuit transpiles to 18 two-qubit gates at depth 22, fewer than PEF-24, but measures more than three times as many qubits.  Its gate-only proxy is therefore higher ($0.96$ versus $0.91$), whereas the gate-plus-readout proxy is lower ($0.57$ versus $0.82$).  This illustrates why $\nzz$ is a necessary resource count but not a cross-register performance model; placement and job snapshot remain confounded in the proxy comparison.

\section{Interpretation and scope}
\label{sec:algorithms}
\subsection{Matched sample-space control}
A complementary simulated mixer comparison illustrates why the reference distribution must match the supplied information. A cardinality-preserving QWOA control starts in a Dicke state in the known optimum sector, whereas full-register QAOA does not receive that cardinality prior. After matching the instance, sector, input and optimization protocol, the weighted-to-unit-weight median optimum-probability ratio is $1.02$, with seed-bootstrap interval $[0.95,1.17]$. This control concerns a dense 12-base, $p=1$ family and exact spectral evolution. Its construction and complete results are in Supplemental Sec.~\ref*{sm:sm:same_sector_overview}.

\subsection{Scope of the diagnostic claims}
\label{ssec:analog_qwoa}
\label{ssec:limitations}
The workflow addresses three potential sources of misinterpretation. Encoding checks prevent a low surrogate energy from being mistaken for a minimum cover. Structural certificates supply an ideal reference that distinguishes ansatz-point limitations from measured loss. Matching the sample space and fixed circuits then gives interpretable hardware contrasts. These are sufficient checks, not a classification of classical difficulty: sparse expander families can have linear treewidth \cite{grohe2009tree}, and failure of our certificates is not a hardness proof.

The hardware contrasts remain observational. Joint execution changes idling, scheduling and simultaneous measurement, so joint/slice retention is not a crosstalk rate. Three jobs, two on one day, cannot establish stationarity or identify a calibration cause. The mixer result is restricted to the stated $\nB=12$, $p=1$ family and exact spectral evolution; neither its one-sided cost oracle nor the $XY$ circuit was digitally implemented. A separate negative audit concerns direct vertex-to-atom Rydberg encoding only, not gadget constructions \cite{nguyen2023quantum} [Sec.~\ref*{sm:sm:ssec:rydberg}].

Finally, the synthetic target rules and incomplete Arco-79 selection provenance limit geographic generalization. Persistent surveillance would require vehicle-indexed routes, flow conservation, initial positions, handoffs and endurance constraints. Those variables can couple the static blocks, so operational feasibility and temporal simulability are outside the present model.

\section{Conclusion}
\label{sec:conclusion}
On these coverage benchmarks, encoding checks and exact structural references determine what a hardware observation can resolve. The surrogate's minima do not represent minimum covers. For the faithful circuit, optimum counts are uninformative at the executed angles and shot budget, and energy reduction is largely reproduced by measured marginals. Although the global weighted-covariance ratio is near $0.50$, a low-weight five-variable block shows a resolved deficit relative to that aggregate in all five examined contexts under the adopted conditional counting model. Decomposing the same moment and propagating its joint counting uncertainty identifies a local discrepancy without interpreting it as fidelity or assigning an error mechanism.

The three-job campaign remains descriptive, and E01 drives its observed joint/slice dispersion. Physical localization and recovered job-associated calibrations show that independent readout at the recorded rates alone cannot reproduce the local covariance deficit; the remaining mechanism is unresolved. The transferable procedure is to verify the target encoding, obtain an ideal structural reference, test whether the chosen event has sufficient probability, and inspect marginal, block and between-block summaries at their stated resolution. Its evidence is restricted to these output distributions; it establishes neither operational surveillance performance nor quantum advantage.

\section{Methods}
\label{sec:methods}

\textbf{Geographic data and distances.}  PEF base coordinates are the surveyed positions in the Army registry \cite{MinistDefesa2023}; the municipality sites are the official IBGE 2024 product \cite{IBGE2024AmazoniaLegal}, centroids computed after projection to SIRGAS 2000/Brazil Polyconic (EPSG:5880).  Separations are haversine great-circle distances [Eq.~(\ref{eq:haversine})].  Observation waypoints are synthetic, drawn as described in Sec.~\ref*{sm:sm:details} without restriction to national territory; the Legal Amazon outline of Fig.~\ref{fig:blocks}(a) is the dissolved IBGE 2024 municipality polygons, simplified at $0.005^\circ$.  The frozen inputs, raw counts and hardware archives are included in the versioned code-and-data release (see Data and code availability).

\textbf{Cost operators.}  Two conventions for $m$ appear in this paper and must not be conflated: the \emph{strict} convention, used by the integer programs and the convergence audit, requires $m$ covers of every task; the \emph{clipped} convention, $t_t=\min(m,\deg t)$, is used only for older archived QUBO benchmarks.  QUBO coefficients follow Eq.~(\ref{eq:qubo_coeffs}) with $\lambda=10$ for PEF-24 and Arco-79, and the audits use Supplemental Eq.~(\ref*{sm:eq:qwoa_audit_cost}) with $\lambda=\nB+1=13$ [Sec.~\ref*{sm:sm:details}].

\textbf{Classical solvers.}  Exact optima come from HiGHS through \texttt{scipy.milp} \cite{huangfu2018highs}, with time limits given in \cite{SMV7}; LP bounds are obtained from the continuous relaxation of the same model.  Reported wall times were measured using the solver's default settings on a single x86-64 workstation and should be treated as indicative rather than intrinsic properties of the instances. The conclusions rely instead on exactly reproducible structural quantities: optima, block counts, treewidth bounds, and integrality gaps.

\textbf{Superconducting hardware.}  Hardware data come from eleven \texttt{SamplerV2} jobs on \texttt{ibm\_kingston} (Heron r2, 156 qubits).  Three standardized jobs reuse identical archived ISA circuits without re-transpilation; each contains four randomized rounds, 60 pubs and $6\times10^5$ shots with XY4 dynamical decoupling and gate and measurement Pauli twirling.  All other QAOA jobs, block-wise jobs and the second faithful-encoding job used the same mitigation, while the first faithful job used sampler defaults.  Physical counts, depths and job identifiers are archived. Job-associated gate and readout errors are retained for the historical resource-proxy circuits. For E01--E03, full backend properties were recovered retrospectively through the job records, with job IDs, backend and calibration dates checked against the original archives; parameter timestamps and units are preserved (Supplemental Sec.~\ref*{sm:sm:physical_audit}).  The 24-qubit PEF-24 pair is treated as a fixed pilot parameter. Its missing optimization record limits claims about training; all output diagnostics use the exact reference at that executed pair. The faithful 51-qubit angles were obtained through a separate offline minimization of the exact block-factorized expected energy (Sec.~\ref*{sm:sm:settings}). The Arco-79 pair gives the largest ideal $P(\mathrm{opt})$ found over the tested $p=1$ domain using the known optimum \cite{SMV7}.  Tiled $p=2,3$ angles are depth controls only.

\textbf{Auxiliary controls.} The mixer study matches the cardinality sector, Dicke input, objective and optimization protocol; its generators, angle windows, convergence audit and gradient checks are specified in Supplemental Secs.~\ref*{sm:sm:settings} and \ref*{sm:sm:same_sector_overview}. A separate enumerated Rydberg control tests direct encoding of the target set, as detailed in Supplemental Sec.~\ref*{sm:sm:ssec:rydberg}.

\textbf{Success-proxy model.}  Cross-register comparisons use Eq.~(\ref{eq:resource_exact}) per coupler and measured qubit, with the properties snapshot associated with each job.  The resulting proxies are $0.817$ for PEF-24, $0.573$ for Arco-79 and $0.316$, $0.306$ for the two historical faithful executions.  They omit idling and correlated errors and are resource diagnostics, not calibrated stochastic output models.

\begin{acknowledgments}
This study was financed, in part, by the S\~ao Paulo Research Foundation (FAPESP), Brasil. Process Number \mbox{2026/04387-7} (F.\ S.\ Luiz). M. C. de Oliveira acknowledges financial support from the National Institute of Science and Technology for Applied Quantum Computing through CNPq process No.\ 408884/2024-0, by FAPESP through the Center for Research and Innovation on Smart and Quantum Materials (CRISQuaM), process No.\ 2013/07276-1, and the support by Venturus Innovation Center.
\end{acknowledgments}

\section*{Data and code availability}
Code and reproducibility data are archived as MIT-licensed release 8.1.0 in Zenodo, DOI \href{https://doi.org/10.5281/zenodo.24540635}{10.5281/zenodo.24540635} \cite{Luiz2026QAOADiagnostics}, and at \href{https://github.com/fsluiz/qaoa-coverage-diagnostics/releases/tag/v8.1.0}{github.com/fsluiz/qaoa-coverage-diagnostics, release v8.1.0}. The archive includes frozen instances, raw bitstring counts, native circuit instruction lists and QPY files, physical layouts, job-associated calibrations, supporting numerical archives, a SHA-256 input manifest and offline reproduction instructions. Source attribution for geographic inputs is retained in the data documentation. The article and Supplemental Material are distributed separately through arXiv.

The Supplemental Material also includes Refs.~\cite{jaksch2000fast,wurtz2023aquila,quintanilla2000efficient,mertens2012continuum}.

\FloatBarrier
\clearpage
\section*{Supplemental Material}
\setcounter{section}{0}
\setcounter{subsection}{0}
\setcounter{equation}{0}
\setcounter{figure}{0}
\setcounter{table}{0}
\renewcommand{\thesection}{S\arabic{section}}
\renewcommand{\thesubsection}{S\arabic{section}.\arabic{subsection}}
\renewcommand{\thefigure}{S\arabic{figure}}
\renewcommand{\thetable}{S\arabic{table}}
\renewcommand{\theequation}{S\arabic{equation}}
\makeatletter
\renewcommand{\p@subsection}{}
\makeatother
\providecommand{\theHsection}{}
\renewcommand{\theHsection}{supp.\arabic{section}}
\providecommand{\theHequation}{}
\renewcommand{\theHequation}{supp.\arabic{equation}}
\providecommand{\theHfigure}{}
\renewcommand{\theHfigure}{supp.\arabic{figure}}
\providecommand{\theHtable}{}
\renewcommand{\theHtable}{supp.\arabic{table}}

\section{Instances, conventions and auxiliary analyses}
\label{sm:sm:details}
We distinguish the number of candidate bases $\nB$, reachable targets $\nt$ and circuit
qubits $\nq$.  The surrogate uses $\nq=\nB$; the faithful slack encoding adds target-local
binary registers.  For the ensemble, the length scale $\delta$ is the median nearest-neighbor
distance in a reference subset of $\nB$ of the 773 municipality centroids, drawn without
replacement by a freshly initialized \texttt{default\_rng(0)} for each $\nB$.
It is fixed across the six radii and 25 instances at that size, not recalculated per instance.

\begin{table*}[t]
\caption{Instance families and their distinct roles. Targets are synthetic except in the municipality dominating-set instance. Hardware registers include slack where indicated; the mixer control is simulated spectrally.}
\label{sm:tab:sm_families}
\small
\begin{ruledtabular}
\begin{tabular}{p{0.15\textwidth}p{0.27\textwidth}p{0.13\textwidth}p{0.30\textwidth}}
Family & Candidate bases and targets & Range & Role \\
\hline
PEF-24 & 24 fixed posts; 23 selected waypoints & 200 km & 24q surrogate and 51q faithful hardware references \\
Arco-79 & 79 fixed centroids; 49 reachable targets & 100 km & 79q surrogate and ideal ansatz-point control \\
Legal-Amazon-773 & 773 centroids as bases and targets & 200 km & Constrained optimum and surrogate-width audit \\
Mixer control & One fixed set of 12 posts; 23 targets per seed & 500--720 km & 20 task seeds, strict $m=1,2,3$, $p=1$ \\
Geographic ensemble & 20, 40, 80 or 160 sampled centroids & Six $R/\delta$ values & 600 feasible instances, structural associations \\
Refinements & Frontier, lattice and irregular deployments & Generator-dependent & Width and LP/ILP controls; see SM \\
\end{tabular}
\end{ruledtabular}
\end{table*}

\subsection{Conventions and forced bases}
The forced fraction is $\phi=n_f/\kopt$, where $n_f$ counts bases fixed by iterative unit
propagation of the strict covering constraints.  After each assignment, reduce every target's
unmet demand by its selected incident bases; if that demand equals the number of still
available incident bases, all of them are forced.  Iterate to a fixed point.
For an unmet target, selecting $f_t$ incident bases leaves demand $m-f_t$ and $d_t-f_t$ candidates; the slack $d_t-m$ is invariant. Therefore this rule forces exactly the union of bases incident to initially tight rows ($d_t=m$), with degree-one forcing at $m=1$. Selecting these bases can satisfy other targets but cannot create a new tight row. Cascades would require additional operations, such as fixing candidates to zero, outside this rule.  The procedure gives a sound forced set, not all variables shared by every optimum.

\subsection{Frozen geographic instances}
\textbf{PEF-24 instance selection.}  The bases are surveyed Army-registry positions \cite{MinistDefesa2023}; the synthetic waypoints are drawn by anchoring each to a uniformly chosen base at a random bearing and an offset of at most $0.7R$.  Eleven of 23 fall outside the IBGE boundary, up to 92 km across the border [Fig.~\ref*{fig:blocks}(a)], so they are not operational sitings.  A draw is accepted when every waypoint is reachable and the greedy cover size is ten.  Seed 0 is the first accepted draw.  Thus $\kopt=10$ is a property of a selected frozen instance, and $k_\mathrm{greedy}=\kopt$ is not evidence about greedy in general.  Coordinates, coverage matrix and couplings are archived in \texttt{frozen\_rung1.json}; Arco-79 is frozen separately.

\textbf{Arco-79 site verification.}  We verified each site against the IBGE product: all 79 names match and the coordinates agree with the official centroids to a median of $98$ m, with a single outlier (Nova Ubirat\~a, MT, $22.7$ km) consistent with a boundary revision in a municipality of that size.  Rebuilding the instance from the exact centroids leaves $\nt$, the link count, $\nzz$ and $\kopt$ unchanged.

\subsection{Cost operators and mixer family}
\textbf{Cost-operator and mixer conventions.}  The \emph{strict} convention requires $m$ covers of every task and is used by the integer programs and convergence audit.  The \emph{clipped} convention uses $t_t=\min(m,\deg t)$ only for older archived QUBO benchmarks.  PEF-24 and Arco-79 use Eq.~\ref*{eq:qubo_coeffs} with $\lambda=10$; the mixer audit uses the faithful threshold cost of Eq.~\ref{sm:eq:qwoa_audit_cost} with $\lambda=\nB+1=13$.  Its exact global evolution $\exp[-i\beta H_{XY}(w)]$ is obtained by diagonalization in the fixed-weight sector; no sequential digital $R_{xx}R_{yy}$ circuit is used.

\textbf{Twelve-base mixer family.} The construction in \path{scripts/benchmark_qwoa_qaoa.py} first selects one fixed cluster from \path{data/pef_planilha.xlsx}: for every post, sum distances to its 11 nearest other posts, take the minimizing post and retain its 12-post neighborhood. The same coordinates are used for all task seeds. With \texttt{RandomState(seed)}, draw 23 task latitudes and longitudes independently and uniformly in the cluster's bounding box, expanded by 10\% of each coordinate span on each side. Search $R=80,100,\ldots,1200$ km for the first incidence matrix with minimum target degree at least three and mean degree in $[3.5,6]$. If none exists, search $R=80,100,\ldots,2000$ km for the first radius meeting minimum degree three alone. The alternative rule is used in 16 of the 20 seeds. Actual radii are 500--720 km (median 630), mean target degrees 5.48--8.65, and support edge counts 49--66 (median 59). Four supports are complete. These metadata are reconstructed by \path{revision_v7/mixer_summary.py} without rerunning the optimizations.

Using the main-text convention $A_{tb}$ (the transpose of the code's \texttt{A\_q}), define
\begin{equation}
 \begin{aligned}
 w_{bb'}&=\sum_t A_{tb}A_{tb'},\\
 H_{XY}(w)&=\tfrac12\sum_{b<b':w_{bb'}>0}w_{bb'}(X_bX_{b'}+Y_bY_{b'}).
 \end{aligned}
 \label{sm:eq:sm_xy}
\end{equation}
The off-diagonal swap matrix element in a weight sector is $w_{bb'}$. The generator can bridge disconnected components by weight-0.1 edges between successive minimum-index representatives; no such bridge is activated in these 20 instances. The primary normalized weights divide by $[|E|^{-1}\sum_{e\in E}w_e^2]^{1/2}$; the support control replaces each weight by one and the complete control includes all 66 pairs. Exact sector diagonalization evolves the full sum, rather than a sequence of digital edge rotations.

\subsection{Refinements and optimization certificates}
\textbf{Refinement and complete-coverage controls.} In \path{scripts/calc_dimension_sweep.py}, order the PEF posts by decreasing latitude and interpolate along successive segments: a segment of length $L$ uses $\max(\lfloor L/s\rfloor,1)$ equal coordinate steps at nominal spacing $s$. Frontier bases use $s=R$ and targets use 25 km, for $R=300,200,150,100,75,50$ km. On the centroid bounding box, the lattice uses latitude pitch $R/111$ degrees and longitude pitch $R/[111\max(\cos\theta,0.2)]$; targets are offset by $R/222$ degrees in both coordinates. Its radii are 400,300,200,150,100 km. The irregular arm instead draws independent uniform latitude--longitude coordinates, with $n=\max(\lfloor |\Omega|/R^2\rfloor,10)$ bases and targets and \texttt{default\_rng(3)}, at 400,300,200,150,120 km. Idle bases and unreachable targets are removed in this arm, and the original coverage fraction is recorded. These eleven regular and five under-dense instances form the sixteen-row refinement sweep; only the regular families provide complete-coverage examples. Records are in \path{results/dimension_sweep.json}.

The separate irregular complete-coverage control, \path{scripts/calc_coverage_preserving_2d.py}, redraws bases and targets from a newly initialized generator at each attempt, increasing the base count by $n_B\mapsto\lfloor1.25n_B\rfloor+1$ until every target is covered; idle bases are then removed. Targets number $\max(\lfloor|\Omega|/R^2\rfloor,10)$, with a 40,000-base construction cap. The archive \path{results/coverage_preserving_2d.json} records the radius sweep, seeds 3--5 and the scale extension down to 60 km. Because each attempt reinitializes the generator and changes array lengths, this is a redraw rule, not incremental addition to a fixed target set.

\textbf{LP/ILP certificates and limits.} HiGHS is called through SciPy with binary bounds $[0,1]$ for ILP and their continuous relaxation for LP. The refinement ILPs use a 300-s cap and the complete-coverage controls 1200 s (relative MIP gap zero); the 600-instance ensemble uses 30 s. Reported certified optima require status zero; the corresponding status, LP bound and incumbent are archived. The municipality archive \path{results/ilp_geographic_ibge_2024_centroids.json} records a 60-s limit, status, MIP gap and selected municipality codes. These numerical certificates are distinct from the graph-width upper bounds returned by the min-fill heuristic.

\subsection{Geometric checks and scope}
\textbf{Where blocks survive: under-dense deployments.}  Drawing bases and targets from a uniform
process at fixed $\rho R^2=1$ leaves $4$--$8\%$ of targets unreachable and yields $7$--$20$
components, with mean interaction degree $\langle d_{ZZ}\rangle=2\nzz/\nB=4.98$--$5.60$.  Raising the
base intensity until every target is covered lifts $\langle d_{ZZ}\rangle$ to $8.0$--$21.5$ and merges
the components into one to seven, the largest holding at least $88\%$ of the bases.  The coverage
shortfall and the blocks are two symptoms of one under-dense deployment, not cause and effect.
We do not compare these degrees with a percolation threshold.  $G_{ZZ}$ joins two bases when some
target lies within $R$ of both, so it is an intersection graph of a two-species (base--target)
process, rather than the Poisson monodisperse disk-intersection graph. For disks of radius $r$, the latter has filling factor $\eta=\rho\pi r^2$ and threshold $\eta_c\simeq1.12808737$; connection distance $2r$ gives mean degree $4\eta$ and critical mean degree approximately $4.51235$ \cite{quintanilla2000efficient,mertens2012continuum}. Neither threshold applies directly to our incidence-derived graph.

\textbf{Strip-bound validation.}  On real geography the strip occupancy must be measured using
a 1-Lipschitz coordinate.  We minimize over the north--south signed distance and the signed
distance to a fixed meridian, both great-circle coordinates of the form specified in the main
text.  The worst $2R$ strip across the 773 municipality centroids then holds 228 of them (the
other admissible orientation holds 259).  As a reference we
draw 773 points uniformly (in area on the sphere) inside the official outline
(\texttt{legal\_amazon\_outline\_2024.geojson}, area $5.06\times10^6$ km$^2$ from the acceptance
rate) and apply the same minimum over admissible coordinates: $149$ on average over 200 draws,
$95\%$ range $136$--$165$,
maximum $173$ (\texttt{calc\_ladder\_rungs\_v5.py}).  The real centroids exceed every draw, by a factor
of $1.5$ on average.  An earlier comparator used the mean density over the latitude--longitude
bounding box ($8.7\times10^6$ km$^2$), whose empty area inflated the apparent factor to $1.8$.  On
the lattice sweep ($H=2502$ km, $\rho_{\max}=1/R^2$) the continuum estimate $W\approx2\rho_{\max}RH=2H/R$,
not a measured strip count, gives $12.5,\,16.7,\,25.0,\,33.4,\,50.0$ against measured min-fill
upper bounds $14,\,15,\,24,\,33,\,48$.  The continuum expression is an occupancy estimate, whereas min-fill supplies a graph-width
upper bound; neither is a lower bound.  Their proximity does not validate the asymptotic
hypotheses.  The occupancy coefficient is $2H/\sqrt{|\Omega|}=1.75$, compared with $1.62$
in the two largest min-fill results.

\textbf{Temporal scope.}  Persistent surveillance is outside the model analyzed here.  A faithful formulation requires vehicle-indexed flow or path variables, conservation, initial positions, handoffs and endurance constraints for individual vehicles.  Those additions may couple distinct static blocks, so no temporal feasibility, treewidth or classical-tractability conclusion is inferred from the static benchmark.

\section{Structural identities and simulation bounds}
\label{sm:sm:structure}

\subsection{Encoding and block probabilities}
Each target of degree $d_t$ contributes $\binom{d_t}{2}$ base--base pair terms
before repeated pairs are merged.  The raw sums for PEF-24 and Arco-79 are 40 and
10, but there are only 17 and 9 distinct base--base edges.
For $\ell_t=\lceil\log_2(d_t-m+1)\rceil$ slack bits, the additional raw counts are
$d_t\ell_t+\binom{\ell_t}{2}$.  The faithful totals are therefore 126 and 30 raw
terms, versus 103 and 29 distinct interactions.  Extra binary values above
$d_t-m$ cannot satisfy a slack equality and do not introduce spurious feasible
ground states.  Ground-state faithfulness follows from a sharper repair bound for feasible instances with unit base costs and private slack. For strict demand $m_t\le d_t$, let $r_t=\max\{0,m_t-(Ax)_t\}$. At any slack assignment the penalty is at least $\lambda\sum_t r_t^2$. Adding bases until coverage is feasible requires at most $\sum_t r_t$ additions: each addition reduces at least one positive integer deficit, and selecting all bases is feasible. Resetting every private slack then gives energy at most $|x|+\sum_t r_t$. If any deficit is positive, $\lambda>1$ makes this repair strictly better. If the bases are feasible but slack is inconsistent, correcting slack lowers the penalty without changing cost. Thus every ground state is feasible and minimizes cardinality. A single base covering a single target ties the empty assignment at $\lambda=1$. This guarantee assumes no additional budget, routing or incompatibility constraints. The hardware value remains $\lambda=10$. Independent component enumeration gives ground energy 10 with 288 feasible minimizers for PEF-24 and 31 with four for Arco-79.

For a product input and block-local evolution, let $p_i(j)$ be the probability of
weight $j$ in block $i$.  Then
\begin{equation}
 P(|x|=k)=[z^k]\prod_i\left(\sum_{j=0}^{|V_i|}p_i(j)z^j\right).
 \label{sm:eq:sm_convolution}
\end{equation}
Replacing $p_i(j)$ by the probability of feasible block configurations of weight
$j$ gives the joint feasible-and-weight distribution.  At the minimum cardinality,
feasibility factorizes over components because every target's incident bases lie
in one component.  These are exact identities; state propagation and convolution
add gate- and degree-dependent polynomial work to the represented dimension.

\subsection{Dicke input and block-local number conservation}
\label{sm:sm:dicke}
Let $n_i=|V_i|$, $n=\sum_i n_i$, and $\ket{D_n^k}$ be the uniform normalized
weight-$k$ Dicke state.  Grouping its computational-basis strings by block weights gives
\begin{equation}
 \ket{D_n^k}=
 \sum_{\substack{0\le k_i\le n_i\\\sum_i k_i=k}}
 \sqrt{\frac{\prod_i\binom{n_i}{k_i}}{\binom nk}}\,
 \bigotimes_i\ket{D_{n_i}^{k_i}}.
 \label{sm:eq:sm_dicke_blocks}
\end{equation}
A diagonal block-local cost and an $XY$ mixer supported within the same components
preserve each $k_i$.  Computational-basis outcomes from different allocations are
disjoint, so their measurement distribution is a mixture: first sample the
multivariate hypergeometric allocation with weights
$\prod_i\binom{n_i}{k_i}/\binom nk$, then independently sample the evolved local
states conditional on those weights.  Dynamic programming or sequential
hypergeometric sampling handles the allocation with polynomial overhead.
Thus bounded blocks remain efficiently sampleable despite entanglement in the input.

Across a cut consisting of whole blocks, with $a$ qubits on one side, the squared
Schmidt coefficients of the input are
\begin{equation}
 \lambda_j=\frac{\binom aj\binom{n-a}{k-j}}{\binom nk},
 \quad \max(0,k-n+a)\le j\le\min(k,a).
 \label{sm:eq:sm_dicke_schmidt}
\end{equation}
Evolution factorizes across this cut and leaves this spectrum unchanged.
The rank is at most $k+1$ (more tightly $\min(k,n-k,a,n-a)+1$), not exponential
in the number of blocks.  If a global optimum requires the component-wise
minimum-cardinality allocation $(k_i^\star)$, then at $k=k^\star$
\begin{equation}
 P(\mathrm{opt})\le
 \frac{\prod_i\binom{n_i}{k_i^\star}}{\binom n{k^\star}}.
 \label{sm:eq:sm_dicke_ceiling}
\end{equation}
The bound is input sector mass, not a claim that a chosen mixer attains it.
It does not apply to mixers coupling different components.

\subsection{Physical strips, slack registers and marginal contraction}
\label{sm:sm:contraction}
For any 1-Lipschitz coordinate $\xi$, the bags
$B_s=\{v:s\le\xi(v)\le s+2R\}$ cover each interaction edge and contain each
vertex over one interval of $s$.  Taking the finite sequence at changes of
membership gives a path decomposition of width at most $\max_s|B_s|-1$.
This combinatorial statement needs no density model.

To estimate that occupancy from geometry, more is required: use physical planar
strips of width $2R$ in a deployment of transverse extent $H$, and assume every
radius-$R$ disk contains at most $M_B$ bases.  A strip can be covered by
$O(1+H/R)$ such disks, hence
$W=O(M_B(1+H/R))$.  An arbitrary 1-Lipschitz coordinate need not have this
strip-covering property (a constant coordinate is a counterexample).
For approximately square, quasi-uniform deployments, the additional two-sided
scaling $|\Omega|=\Theta(n/\rho)$, $H=\Theta(\sqrt{|\Omega|})$ and
$\rho R^2=\Theta(1)$ with $M_B=O(1)$ yields $W=O(\sqrt n)$.
An upper density bound alone does not imply the area relation.
These are asymptotic hypotheses, not conclusions from the finite Amazon map.

Place slack bits at their target's coordinate.  Base--slack edges have length at
most $R$, slack--slack edges within a target have length zero, and base--base
edges have length at most $2R$.  The same construction gives
$w\le W_B+W_S-1$, where $W_B,W_S$ are the maximum base and slack occupancies.
To control $W_S$, assign each target to one covering base.  A target in a $2R$
strip has its assigned base in the expanded $4R$ strip, covered by two base bags.
If each base covers at most $M_T$ targets and every target has at most $L$ slack
bits, $W_S\le2M_TLW_B$.  Bounded $M_T,L$ therefore preserve the order of the
width bound.  Without a target-density or degree bound, this conclusion need
not hold for the faithful encoding.

For product-input, diagonal two-body QAOA with one-qubit mixers, introduce one
binary variable per wire and layer.  An amplitude is a sum over these variables;
a probability or marginal requires both a ket and a bra copy, with final-index
identifications and projectors on the observed qubits.  From a decomposition
of the spatial interaction graph with bags $X$ of size at most $w+1$, form
\begin{equation}
 \widetilde X=X\times\{0,\ldots,p\}\times\{\mathrm{ket},\mathrm{bra}\}.
\end{equation}
Every cost factor, mixer factor and final identification lies within one bag;
the running-intersection property follows from that of $X$.
The bag size is at most $2(p+1)(w+1)$.  Standard variable elimination
therefore evaluates a marginal in
$\operatorname{poly}(N_g)\,2^{O((p+1)(w+1))}$ arithmetic operations.
Fixing successive measured bits and evaluating their conditional probabilities
uses at most a linear number of these contractions to sample a full outcome;
branches of zero probability are not selected.

This is an upper bound, not a matching lower bound or a uniform finite-precision
bit-complexity guarantee.  It does not cover an arbitrary nonlocal mixer.
For $w=O(\sqrt n)$ the bound is subexponential at fixed depth, and remains so
when $p=o(\sqrt n)$; unrestricted growing depth is not covered.

\subsection{Connected graphs and width certificates}
\label{sm:sm:width_certificates}
Complete coverage does not imply a connected support graph: separate groups can cover disjoint target sets. Connectivity holds in our refinement constructions because the base-overlap graph is connected and every required overlap contains a target. Keeping coverage complete while reducing $R$ gives connected instances along the frontier ($\nB=35$--197) and on a Legal Amazon lattice ($\nB=78$--879). The under-dense controls in Sec.~\ref{sm:sm:details} show how fragmentation depends on the chosen deployment, not on coverage constraints alone.

In two dimensions exact blocks give way to a width bound.  Let $\xi:B\to\mathbb R$ be a 1-Lipschitz coordinate, $|\xi(b)-\xi(b')|\le d(b,b')$, and define $\mathcal B_s=\{b:\xi(b)\in[s,s+2R]\}$.  If $(b,b')\in E(G_{ZZ})$, then $|\xi(b)-\xi(b')|\le2R$, so some bag contains both endpoints; the bags containing each vertex form an interval.  Retaining endpoints at vertex coordinates gives a finite path decomposition and
\begin{equation}
  \label{sm:eq:sm_strip}
  \mathrm{tw}(G_{ZZ})\le\mathrm{pw}(G_{ZZ})\le W_\xi-1,
  \qquad W_\xi=\max_s|\mathcal B_s|.
\end{equation}
The north--south coordinate $\xi_{\rm NS}(b)=r_\oplus(\theta_b-\theta_0)$ is 1-Lipschitz under great-circle distance.  More generally, signed distance to a fixed great circle, $\xi_{\boldsymbol n}(b)=r_\oplus\arcsin(\boldsymbol n\!\cdot\!\boldsymbol u_b)$ for site unit vector $\boldsymbol u_b$, gives valid sweeps at arbitrary orientations.  An arbitrary map projection does not suffice.

To convert occupancy into an asymptotic bound, additional geometric assumptions are needed. For Cartesian strips in a planar domain of transverse extent $H$, a local cap of $O(\rho R^2)$ bases per radius-$R$ disk gives
$W=O[\rho R^2(1+H/R)]$; being 1-Lipschitz alone does not guarantee such physical strips. At fixed local occupancy, straight corridors with $H=O(R)$ have bounded width. In a bounded-aspect, quasi-uniform planar family with area $\Theta(\nB/\rho)$ and $\rho R^2=\Theta(1)$, the bound is $O(\sqrt{\nB})$. The relation between area and density is an assumption, not a consequence of an upper density cap.

For the slack encoding, assigning each target's private bits its coordinate gives $\mathrm{pw}(G_{ZZ}^{\rm slack})\le W_B+W_S-1$. The same scaling follows if every radius-$R$ disk contains a bounded number of targets and target degree is bounded; Sec.~\ref{sm:sm:contraction} gives the explicit bound and its scope.

On the 773 centroids the smaller maximum occupancy from two admissible signed great-circle coordinates is 228. Uniform draws inside the official IBGE outline give 149 on average (95\% range 136--165; maximum 173 in 200 draws), about a factor 1.5 lower. The sampled frontier refinements have measured min-fill width bounds decreasing from 10 to 6 as $\nB$ grows by a factor 5.6; these observations do not establish a theorem for arbitrary curved corridors. Geometric separators provide related bounds under a bounded-ply hypothesis \cite{miller1997separators}.

The simulation consequence concerns the circuit, not the encoded optimization problem. For a diagonal QAOA cost and one-qubit mixers, lifting a width-$w$ interaction-graph decomposition through $p+1$ layers gives amplitude bags of size at most $(p+1)(w+1)$. A doubled bra--ket construction for marginal probabilities has bags of size at most $2(p+1)(w+1)$. Hence contraction and sequential conditional sampling require
\begin{equation}
 \label{sm:eq:sm_contraction}
 N_g^{O(1)}2^{O((p+1)(w+1))}
\end{equation}
arithmetic operations \cite{markov2008simulating}; the construction and precision convention are in Sec.~\ref{sm:sm:contraction}. For fixed $p$, the deterministic corridor and planar regimes above give polynomial and subexponential bounds, respectively. For growing depth the planar exponent is sublinear only when $p=o(\sqrt{\nB})$.

These certificates need not explain fast optimization. The connected $\nB=879$ instance has width upper bound 48 yet its ILP closes in under 50 ms. Across the sixteen refinement instances, the rounded-up LP bound equals the integer optimum; irregular complete-coverage controls have some nonzero gaps but also solve quickly. These are observations about the tested generators, not general tractability claims [Sec.~\ref{sm:sm:details}].

The 773-base bare-register surrogate marks the limit of these sufficient certificates.  Its interaction graph has one component of 770 vertices and three isolated vertices, contains a verified 113-vertex clique and admits a verified min-fill decomposition of width 151.  Thus $112\le\operatorname{tw}\le151$: exact block enumeration is ineffective and the low-width certificate used above is absent.  The graph also has $39{,}974$ logical $ZZ$ interactions before routing.  These facts are resource statements about the two-sided surrogate, not a runtime lower bound or a faithful dominating-set circuit.

\section{Numerical settings}
\label{sm:sm:settings}

\textbf{Angle fixing for the hardware circuits.}  All angles were held fixed on hardware.  At PEF-24 (24 qubits), the archived $p=1$ pair is $(-0.04,0.35)$.  The repository contains the pair but not the search program, objective, seed or output that originally produced it, so we treat it as a fixed pilot parameter and do not claim that it is variationally optimal.  The 8192-shot transferred-angle Arco-79 job uses that same pair.  For the own-angle $10^5$-shot Arco-79 job, block factorization makes the ideal target-set probability exactly computable, and an oracle-assisted search maximizing $P(\mathrm{opt})$ against the known optimum gives $(\gamma,\beta)=(-0.7596125,0.7145860)$.  This is an ansatz-reachability control, not conventional energy training.  Its $p=2,3$ depth controls tile the $p=1$ pair rather than re-optimizing every layer.

The faithful 51-qubit angles were obtained through a separate offline minimization of the exact block-factorized expected energy.

\textbf{Convergence-first QWOA/QAOA audit.}  Two seeds, 12 and 17, were drawn without
replacement before evaluation from a NumPy generator seeded by 20260915 and paired across
strict $m=1,2,3$ at $\nB=12$.  The diagonal objective is
$|x|+13\sum_t\mathbf 1[(Ax)_t<m]$, and both ans{\"a}tze minimize its exact expectation;
$P(\mathrm{opt})$ is not exposed to the optimizer.  The $p=1$ landscape is evaluated on a $41\times41$ grid (cost-angle endpoint excluded, mixer endpoints included) and audited
by simplicial homology global optimization (SHGO) with Sobol sample sizes 1024 and 2048 and iteration counts one and two.  The two
discordant QWOA cases are rerun at 4096 points and three iterations, reproducing one of the
two original endpoints in each case.  For $p=2,3$, three paired differential-evolution searches use 1024
and 2048 objective calls, respectively, before exact-adjoint-gradient L-BFGS-B refinement
(\texttt{maxiter} 400, \texttt{maxfun} 1200, \texttt{ftol} $10^{-13}$, \texttt{gtol} $10^{-9}$).  Identity-layer and repeated-layer extensions
of the best shallower result prevent a spurious loss with depth.  A point is called converged
only when at least two global endpoints lie within $10^{-3}$ in expected energy, at least two
thirds of endpoints meet that tolerance, and their relative $P(\mathrm{opt})$ range is at
most $2\%$.  All six pairs pass at $p=1$ after the adaptive audit; none passes at $p=2$ or 3.  The fixed-sector QWOA phase uses its fundamental $2\pi/13$
cost-angle period; QAOA uses $2\pi$, and both mixer-time searches are restricted to $[0,\pi]$.  The twenty-seed extension (Sec.~\ref{sm:sm:qwoa20}) runs the $p=1$ protocol only, with the
adaptive third resolution applied automatically wherever the first two disagree.

\textbf{Same-sector mixer control.}  The six converged $p=1$ cases are rerun from the same
$k^\star$ Dicke state with the same exact cost, parameter bounds and global-search protocol.
The primary unit-weight control preserves the weighted cooperation graph's edge support; the
secondary control completes the graph.  A scale sensitivity divides the shared-task weights
by their edge-weight RMS.  Every new endpoint passes the joint convergence criterion, and
the adjoint gradients for all three control families match central differences to
$2.3\times10^{-9}$ or better.

\textbf{Rydberg encoding audit.}  We enumerate all 512 configurations of the $3\times3$
grid, its blockade graph and the final classical Hamiltonian.  DOP853 with
\texttt{rtol} $10^{-9}$ and \texttt{atol} $10^{-11}$ reproduces the archived
trajectory; it is used only to distinguish transient target-set overlap from a ground-state
encoding and is not a hardware model.

\textbf{Distances.}  Haversine great-circle distances with $r_\oplus=6371.0$ km for the
frontier instances and $6371.0088$ km in the Legal Amazon ILP; the difference is below
$1$ m at the ranges considered.

\section{Controlled output distributions and diagnostic scope}
\label{sm:sm:diagnostic_controls}
For each component of the faithful 51-variable operator, reconstruct the diagonal energies directly from the slack equalities, apply its executed $p=1$ angles and evaluate first and second bit moments. The largest component has 13 variables and the combined represented dimension is 8958. Distinct components have product moments ideally. This computes $E^{\rm id}=209.1834$, $E_{\rm prod}^{\rm id}=238.9836$ and denominator $29.8002$ without constructing a $2^{51}$ array. \path{revision_v7/diagnostic_controls.py} performs this calculation; it is independently checked against the archived ideal reference.

\begin{figure*}[t]
\centering
\includegraphics[width=\linewidth]{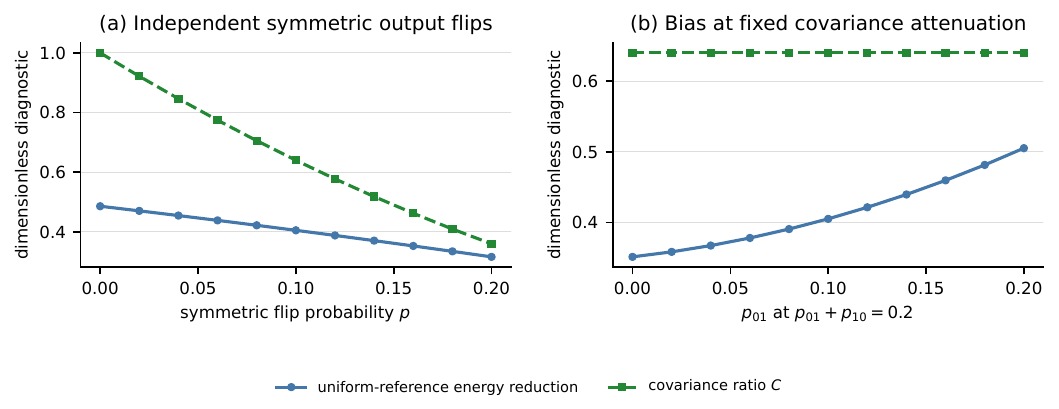}
\caption{Exact synthetic output-channel controls on the faithful ideal reference. Blue: reduction relative to uniform-bit energy; green: weighted-covariance ratio $C$. (a) Independent symmetric bit flips attenuate covariances. (b) Varying bias at fixed $p_{01}+p_{10}=0.2$ changes raw energy reduction while preserving $C=0.64$. The curves illustrate complementary diagnostics under the declared stochastic channels, without identifying a hardware mechanism. The curves are distinct dimensionless diagnostics; their crossings do not define a physical threshold.}
\label{sm:fig:sm_controls}
\end{figure*}

An independent stochastic output channel has $a_i=p_{01,i}$ and $b_i=1-p_{01,i}-p_{10,i}$. For $i\ne j$,
\begin{equation}
 \mu'_i=a_i+b_i\mu_i,\qquad
 \operatorname{Cov}(x'_i,x'_j)=b_ib_j\operatorname{Cov}(x_i,x_j).
 \label{sm:eq:sm_output_channel}
\end{equation}
Diagonal second moments remain $\langle x_i'^2\rangle=\mu'_i$. We declare two grids before calculation: symmetric $p_{01}=p_{10}=0,0.02,\ldots,0.20$, and $p_{01}=0,0.02,\ldots,0.20$ at fixed $p_{01}+p_{10}=0.20$. Symmetric flips give $C=(1-2p)^2$. At fixed transition sum the off-diagonal covariances all scale by $0.64$, while bias changes marginal energy contributions. Figure~\ref{sm:fig:sm_controls} displays the resulting $\ControlBiasReductionMin$--$\ControlBiasReductionMax\%$ range of raw energy reduction at unchanged $C$. The actual flip probability for bit $i$ is $\epsilon_i=(1-\mu_i)p_{01}+\mu_i p_{10}$, so this sweep does not hold its physical error rate fixed. The mean of these probabilities is also stored in the control JSON. These exact output channels specify controlled distributions; they are not a fit to gate, readout or calibration errors of Kingston.

A separate four-bit example establishes a blind spot for block-additive costs. Use $E=-x_0x_1-x_2x_3$ and two blocks $\{0,1\}$ and $\{2,3\}$, each assigning probability $1/2$ to 00 and 11. Their independent product has probability $1/4$ on 0000,0011,1100,1111. A coupled distribution instead assigns $1/2$ to each of 0000 and 1111. Both have $E=-1$, $E_{\rm prod}=-1/2$ and $C=1$ relative to the independent reference. The global optimum probability is respectively $1/4$ and $1/2$, while the product of block-optimum probabilities is $1/4$ in both: the between-block factor changes from one to two. This generic block-additive example is not a new coverage instance. It shows why covariance retention cannot replace the separate dependence comparison.

\section{Device properties and the resource model}
\label{sm:sm:calibration}
The resource model of the main text uses the backend-properties snapshot associated with each
job record, not a single device-wide median.  For every submitted circuit we recover the measured physical
qubits and the multiplicity of each physical two-qubit edge, then evaluate
$\prod_g(1-\epsilon_g)\prod_q(1-\epsilon_{\mathrm{ro},q})$.  The PEF-24 $p=1$ circuit
gives $0.9122\times0.8953=0.8167$ (49 two-qubit gates on 24 measured qubits); the own-angle
Arco-79 circuit gives $0.9642\times0.5945=0.5731$ (18 gates, 79 qubits).  Thus gates alone
favor Arco-79 while readout reverses the ordering.  The transferred-angle job gives
$0.5930$ under its own job-associated snapshot, and the 51-qubit faithful circuit gives
$0.4482\times0.7049=0.3159$ (sampler defaults) and $0.4692\times0.6521=0.3060$ (matched
XY4 and twirling).  The full physical-qubit lists, edge multiplicities, error values,
job dates and properties-snapshot dates are archived in \texttt{resource\_model\_exact.json}; the
calculation is reproduced by \texttt{calc\_resource\_exact.py}.
This is a first-order resource proxy, not a device-noise model.  It omits one-qubit-gate and
reset errors, errors in the dynamical-decoupling pulses, idle-time decoherence, leakage,
crosstalk and other correlated noise.  Consequently percentages attributed to register size,
placement or snapshot-associated device properties are decompositions within this proxy only.
For the standardized three-job campaign below, each job is labeled by the update date of the backend properties
returned by \texttt{job.properties()} after completion. The initial count archives saved only this date; full numerical properties were recovered retrospectively for the physical audit of Sec.~\ref{sm:sm:physical_audit}. In E03 its
\texttt{last\_update\_date} changed from 18:22:17 at submission to 18:25:47 before the recorded
job start on October 14, 2026; the latter timestamp is used in the descriptive
between-job calculation.  The timestamp does not identify which physical parameter changed.

\begin{table*}[t]
  \caption{Principal historical Kingston executions ($10^5$ shots each).  $\Fest$ is the circuit-resource proxy of Eq.~(\ref*{eq:resource_exact}), not a measured fidelity; $\eta_{\rm all}$ uses the unconditional $288/2^{24}$ null.  Energy reductions use encoding-specific random-bitstring references and are not comparable across rows.  Both faithful circuits encode 24 bases and 27 slack bits.}
  \label{sm:tab:kingston}
  \begin{ruledtabular}
    \begin{tabular}{cccccccc}
      Circuit & $\nq$ & $\nzz$ & $n_{2Q}$ & Depth & $\Fest$ & Minimum-cover hits ($\eta_{\rm all}$) & Energy reduction \\
      \hline
      PEF surrogate (24q), $p=1$ & 24 & 17 &  49 &  64 & 0.82 & $79/10^5$ ($\mathbf{46.02}$) & 31.9\% \\
      Arco surrogate (79q), $p=1$ & 79 &  9 &  18 &  22 & 0.57 & $0$ (---) & 32.0\% \\
      Faithful encoding (51q), $p=1$, defaults & 51 & 103 & 372 & 252 & 0.32 & $0$ (---) & 39.7\% \\
      Faithful encoding (51q), $p=1$, XY4+twirling & 51 & 103 & 372 & 251 & 0.31 & $0$ (---) & 36.6\% \\
    \end{tabular}
  \end{ruledtabular}
\end{table*}

\section{Hardware data}
\label{sm:sm:data}

Every hardware number in the main text and this Supplemental Material is recomputed offline
from the following archived files; no IBM Quantum access is required.
The paths in Table~\ref{sm:tab:sm_provenance} are relative to the root of the public code-and-data release. Its \path{PATH_MAPPING.json} records the relocation of historical \path{paper/} analysis paths to \path{analysis/}, while original input bytes and their checksums are preserved. Plans and circuit records retain job identifiers, shots and placements. The historical resource-proxy archive retains job-associated gate and readout errors. Full E01--E03 job properties are included under \path{analysis/revision_v8/calibration_archives/} after retrospective retrieval, preserving the original plans and counts. Section~\ref{sm:sm:physical_audit} distinguishes the available inputs.
\begin{table*}[t]
\caption{Hardware provenance and associated structural archives. All paths are relative to the project root. The README distinguishes the revised reproduction workflow from separately executable historical analyses.}
\label{sm:tab:sm_provenance}
\footnotesize
\begin{ruledtabular}
% Boxed cells avoid the REVTeX/array paragraph-column compatibility issue.
\begin{tabular}{ll}
\parbox[t]{0.61\textwidth}{\raggedright\strut Archive\strut} & \parbox[t]{0.31\textwidth}{\raggedright\strut Role and execution\strut} \\[2pt]
\hline
\parbox[t]{0.61\textwidth}{\raggedright\strut \path{results/kingston_100k_rung1_counts.json}\strut} & \parbox[t]{0.31\textwidth}{\raggedright\strut PEF-24, three depths, $10^5$ shots each; \texttt{dakjch39k43c73aik9rg}\strut} \\[2pt]
\parbox[t]{0.61\textwidth}{\raggedright\strut \path{results/kingston_100k_arco_optangles_counts.json}\strut} & \parbox[t]{0.31\textwidth}{\raggedright\strut Own-angle Arco-79, three depths; \texttt{dakjbtfi3e6s738rm2h0}\strut} \\[2pt]
\parbox[t]{0.61\textwidth}{\raggedright\strut \path{results/kingston_transferred_arco_counts.json}\strut} & \parbox[t]{0.31\textwidth}{\raggedright\strut Transferred-angle circuit 3, 8192 shots; \texttt{d9211s357qjs73b76k00}\strut} \\[2pt]
\parbox[t]{0.61\textwidth}{\raggedright\strut \path{results/exact51_counts.json};\newline\path{results/exact51_matched_counts.json}\strut} & \parbox[t]{0.31\textwidth}{\raggedright\strut Faithful runs; \texttt{dakjjpfi3e6s738rmbng}, \texttt{dat665lvr3kc73el167g}\strut} \\[2pt]
\parbox[t]{0.61\textwidth}{\raggedright\strut \path{results/ibm_block_results/kingston_blocks_analysis.json}\strut} & \parbox[t]{0.31\textwidth}{\raggedright\strut Separately compiled block counts; \texttt{dan62g5r85ps73fdlp00}\strut} \\[2pt]
\parbox[t]{0.61\textwidth}{\raggedright\strut \path{results/joint_separate_batch_counts.json}\strut} & \parbox[t]{0.31\textwidth}{\raggedright\strut Joint and 13 slices with associated snapshot; \texttt{dau04mahcrkc73dur170}\strut} \\[2pt]
\parbox[t]{0.61\textwidth}{\raggedright\strut \path{results/interleaved_batch_plan.json};\newline\path{results/interleaved_batch_counts.json}\strut} & \parbox[t]{0.31\textwidth}{\raggedright\strut 136 pubs, order, gate lists and snapshot ($T_1,T_2$); \texttt{daul9oqhcrkc73dvsqj0}\strut} \\[2pt]
\parbox[t]{0.61\textwidth}{\raggedright\strut \path{analysis/ibm_variance_campaign/results/}\strut} & \parbox[t]{0.31\textwidth}{\raggedright\strut E01--E03 plans, aggregate counts, timestamps and historical analyses\strut} \\[2pt]
\parbox[t]{0.61\textwidth}{\raggedright\strut \path{analysis/revision_v7/results/}\strut} & \parbox[t]{0.31\textwidth}{\raggedright\strut Revised epoch analyses and \path{between_jobs.json}\strut} \\[2pt]
\parbox[t]{0.61\textwidth}{\raggedright\strut \path{results/ibm_circuits/}\strut} & \parbox[t]{0.31\textwidth}{\raggedright\strut Submitted QPY and instruction lists; SHA-256 registry \path{index.json}\strut} \\[2pt]
\parbox[t]{0.61\textwidth}{\raggedright\strut \path{results/rung1_physical_layout.json};\newline\path{results/exact51_physical_layout.json}\strut} & \parbox[t]{0.31\textwidth}{\raggedright\strut Executed layouts and coupling map, retrieved by the respective \path{fetch_*_layout.py} scripts\strut} \\[2pt]
\parbox[t]{0.61\textwidth}{\raggedright\strut \path{results/resource_model_exact.json}\strut} & \parbox[t]{0.31\textwidth}{\raggedright\strut Job-associated properties and physical gate multiplicities\strut} \\[2pt]
\parbox[t]{0.61\textwidth}{\raggedright\strut \path{results/qwoa_simulability_v5.json};\newline\path{results/ladder_rungs_v5.json};\newline\path{results/interleaved_batch_v5.json}\strut} & \parbox[t]{0.31\textwidth}{\raggedright\strut Dicke spectra, admissible strips and round-level interleaved inference\strut} \\[2pt]
\end{tabular}
\end{ruledtabular}
\end{table*}
The analyses are \texttt{analyze\_kingston.py} (QAOA aggregates), \texttt{analyze\_exact51.py}
(faithful-encoding energies and feasibility), \texttt{calc\_statistical\_audit.py} (estimands
and intervals), \texttt{calc\_hardware\_decomposition.py} (sector decomposition and energy
nulls), \texttt{calc\_block\_product\_null.py} (block-product null),
\texttt{calc\_joint\_separate\_batch.py} (batch comparison),
\texttt{calc\_interleaved\_batch\_v5.py} (round-level interleaved analysis), and
\texttt{calc\_exact51\_cross\_block.py} (routing and cross-block correlation).  The three-job analysis uses \path{analysis/revision_v7/analyze_epoch.py},
\path{combine_epochs.py} and \path{campaign_artifacts.py} in the same directory.
The public release uses \path{scripts/reproduce_diagnostics.py} with
\texttt{--output DIR --full}. It verifies \path{DATA_MANIFEST.json}, reruns
the scientific checks, repeats the block and campaign bootstraps, checks the
cross-block correlation points directly from counts, and regenerates the figures
in the chosen output directory. The article sources are unnecessary for this
workflow. Historical manuscript-layout checks remain separate.

\subsection{Physical placement and calibration-data audit}
\label{sm:sm:physical_audit}
The offline audit in \path{analysis/revision_v8/audit_physical_data.py} joins the six interacting logical blocks to the native instruction lists, the measurement maps, the QPY checksum registry, the E01--E03 submission plans and counts, and \path{results/resource_model_exact.json}. It verifies the archived circuit hashes, agreement between instruction-level measurement maps and saved layouts, native two-qubit gate multiplicities, and the four faithful-circuit pubs of $10^4$ shots in every campaign job. Its machine-readable output records the SHA-256 digest of each input in \path{analysis/revision_v8/generated/physical_data_audit.json}.

For each block, the audit starts from its measured physical wires and traverses the native gates backwards, enlarging the wire set whenever an earlier gate touches it. Barriers and measurement instructions do not propagate this influence. The resulting conservative backward cones follow routing through the native gates; their two-qubit instruction counts can overlap and need not sum to the circuit total. This distinction is necessary because a measurement-time placement alone does not assign every routed gate to a unique logical block.

The five-variable block has logical indices $\{0,1,31,32,33\}$. It is measured on physical qubits $\{57,68,70,67,69\}$ in the historical defaults execution and on $\{48,38,30,49,29\}$ in the historical matched execution. The latter QPY file is reused with the same checksum in all four rounds of each E01--E03 job. Table~\ref{sm:tab:sm_physical_campaign} reports all six interacting blocks, rather than selecting the five-variable block alone.

Job-associated gate-error values and readout errors are retained for the two historical executions. Table~\ref{sm:tab:sm_physical_historical} summarizes these by block; the JSON retains the physical indices, gate couplers, shared-gate counts and available per-qubit readout errors. These are descriptive resource quantities. The covariance ratio depends on the signed cost coefficients and joint output distribution, so a calibration-error product does not predict its value without an additional validated channel model.

The initial E01--E03 plans and count archives saved calibration dates but not the error values. The read-only retrieval script \path{analysis/revision_v8/recover_campaign_properties.py} subsequently obtained the backend properties associated with each completed job through \texttt{RuntimeJobV2.properties(refresh=True)}. The three complete responses are saved under \path{analysis/revision_v8/calibration_archives/}, with SHA-256 checksums, retrieval time and SDK version. No new circuits were submitted. Job IDs and backend names agree with the plans, and the returned update dates agree with the count archives: October 8 at 08:55:19, October 14 at 14:27:28, and October 14 at 18:25:47 (UTC$-3$). The parameter-level timestamps and units are retained rather than treating the update date as a simultaneous measurement of every parameter. The historical September errors are not transferred to the October jobs.

Native gate-error records are selected by gate name and physical operands; reversed operands are accepted only for the symmetric CZ gate used here. Every used coupler and all 51 measured qubits have calibration entries. The campaign table reports the gate-count-weighted mean error in each backward cone and the mean and maximum readout error over that block's measured qubits. In notation matching the main text,
\begin{equation}
 \bar\epsilon_{2Q,\alpha,e}
 =\frac{\sum_g m_{\alpha,g}\epsilon_{g,e}}{\sum_gm_{\alpha,g}},
 \qquad
 \bar\epsilon_{{\rm ro},\alpha,e}
 =\frac{1}{|V_\alpha|}\sum_{q\in V_\alpha}\epsilon_{{\rm ro},q,e}.
 \label{sm:eq:sm_local_calibration}
\end{equation}
Here $m_{\alpha,g}$ counts native gate occurrences in the cone and the readout sum uses the block's physical measurement map. These averages summarize the available calibration records; they are not error probabilities estimated from the circuit's bitstring counts.

For the five-variable block, the mean gate errors are $0.2257\%$, $0.2291\%$ and $0.2291\%$ in E01--E03; mean readout errors are $1.5381\%$, $0.9473\%$ and $0.7715\%$. In E01 this block has the largest mean readout error among the six interacting blocks, with a $4.6631\%$ error on physical qubit 30. That qubit's recorded error falls to $0.8545\%$ and $0.7324\%$. This identifies a concrete calibration difference to consider, but it does not explain the full pattern: mean readout error decreases from E02 to E03 while the point covariance ratio also decreases, and the block does not have the largest gate-error average in any of the three jobs. The recorded CZ error values and their parameter timestamps on all 40 used couplers are identical between E02 and E03 despite the different overall properties-update timestamps. Point-value comparisons must also be read alongside the counting intervals in Sec.~\ref{sm:sm:block_profiles}; no between-job difference is declared resolved from this ordering.

\textbf{Independent readout control.} For each measured physical qubit, let $p_{01}=P(1\mid0)$ and $p_{10}=P(0\mid1)$ be the archived directional readout probabilities. Independent binary output channels give, for $i\ne j$,
\begin{equation}
 \operatorname{Cov}(Y_i,Y_j)=b_i b_j\operatorname{Cov}(X_i,X_j),\qquad b_i=1-p_{01,i}-p_{10,i}.
 \label{sm:eq:sm_readout_covariance}
\end{equation}
We apply the individual channels to the exact block distribution at the executed angles rather than replacing them with an average error rate. For the five-variable block, multiplication of the full 32-state distribution by the product transition matrix independently verifies the transformed first and second moments. The asymmetric channels and their symmetrized counterparts under independent measurement twirling have the same $b_i$ and hence the same covariance prediction. The calculation retains every signed QUBO coefficient; it does not assume that opposite-sign couplings cancel. The resulting ratios are $\ReadoutCOne$, $\ReadoutCTwo$ and $\ReadoutCThree$ for E01--E03 (Table~\ref{sm:tab:sm_readout_control}). The predicted attenuation corresponds to $\ReadoutFractionMin$--$\ReadoutFractionMax\%$ of the observed deficit from $C=1$. This is a comparison of conditional point values, not an uncertainty-qualified upper bound or an attribution of a fraction of physical error. It omits correlated readout and calibration drift. Independent readout at the recorded rates alone is insufficient to reproduce the local deficit.

The script \path{analysis/revision_v8/readout_audit.py} records all six interacting blocks, the per-qubit directional rates and their timestamps, input hashes, and the explicit 32-state verification in \path{analysis/revision_v8/generated/readout_audit.json}. A depolarizing model for the 144 native CZ instructions in the five-variable block's backward cone would additionally require a specified conversion from gate-calibration errors to channel parameters and propagation through the routed circuit, including shared wires. Such a model is not inferred from a gate-error average.

\begin{table}[t]
\caption{Independent readout applied to the exact five-variable block distribution. The last column is $(1-C_\alpha^{\rm ro})/(1-C_\alpha^{\rm obs})$, using $C=1$ as ideal reference. Rates are fixed at their archived calibration values; these are model-conditional point comparisons, not causal fractions or confidence bounds.}
\label{sm:tab:sm_readout_control}
\begin{ruledtabular}
\begin{tabular}{lrrr}
Job & $C_\alpha^{\rm obs}$ & $C_\alpha^{\rm ro}$ & Deficit fraction (\%) \\
\hline
E01 & 0.1163 & 0.9549 & 5.11 \\
E02 & 0.2845 & 0.9616 & 5.37 \\
E03 & 0.2304 & 0.9715 & 3.70 \\
\end{tabular}
\end{ruledtabular}
\end{table}

The recovered qubit properties also include $T_1$, $T_2$ and readout durations, with their timestamps. Execution-time schedules, idle durations and an independently justified channel model would still be required to attribute the covariance deficit to a mechanism. The present evidence establishes physical placement, calibration context and the insufficiency of the stated independent readout model. It does not establish that the deficit follows a physical qubit across placements: logical structure, compilation, date and settings change together. The limited campaign does not support a calibrated regression or causal attribution. No calibration value is imputed.

\begin{table*}[t]
\caption{Physical placement and output ratios for all six interacting blocks in E01--E03. Logical block labels refer to the sorted block sizes. Physical qubits are listed in logical-index order. Gate counts include native two-qubit instructions in a conservative backward cone, including shared instructions; counts do not partition the circuit. All three jobs use the same QPY checksum.}
\label{sm:tab:sm_physical_campaign}
\small
\begin{ruledtabular}
\begin{tabular}{rp{0.40\textwidth}rrrr}
Size & Measured physical qubits & Cone gates & $C_{\rm E01}$ & $C_{\rm E02}$ & $C_{\rm E03}$ \\
\hline
4 & 129, 109, 110, 118 & 14 & 0.307 & 0.439 & 0.363 \\
5 & 48, 38, 30, 49, 29 & 144 & 0.116 & 0.285 & 0.230 \\
6 & 97, 107, 104, 117, 106, 105 & 46 & 0.503 & 0.520 & 0.489 \\
7 & 148, 147, 151, 138, 152, 149, 150 & 55 & 0.502 & 0.495 & 0.505 \\
9 & 53, 58, 50, 66, 67, 57, 47, 52, 51 & 158 & 0.539 & 0.511 & 0.544 \\
13 & 41, 36, 23, 68, 46, 44, 45, 26, 25, 24, 37, 21, 22 & 162 & 0.523 & 0.510 & 0.505 \\
\end{tabular}
\end{ruledtabular}
\end{table*}
\begin{table*}[t]
\caption{Recovered job-associated calibration descriptors for all six interacting blocks in E01--E03. Percentages summarize the recorded native CZ errors (gate-count-weighted over the backward cone) and readout errors (mean and maximum over measured block qubits). Parameter timestamps and units are retained in the raw property archives. These descriptors are not fitted error rates for the covariance ratio.}
\label{sm:tab:sm_physical_campaign_errors}
\small
\begin{ruledtabular}
\begin{tabular}{lrrrrr}
Job & Size & $C_\alpha$ & Mean gate error (\%) & Mean readout error (\%) & Max readout error (\%) \\
\hline
E01 & 4 & 0.307 & 0.191 & 0.403 & 0.500 \\
E01 & 5 & 0.116 & 0.226 & 1.538 & 4.663 \\
E01 & 6 & 0.503 & 0.165 & 1.288 & 3.247 \\
E01 & 7 & 0.502 & 0.228 & 0.534 & 0.818 \\
E01 & 9 & 0.539 & 0.215 & 1.154 & 2.124 \\
E01 & 13 & 0.523 & 0.189 & 0.924 & 2.002 \\
E02 & 4 & 0.439 & 0.189 & 0.928 & 1.855 \\
E02 & 5 & 0.285 & 0.229 & 0.947 & 1.270 \\
E02 & 6 & 0.520 & 0.183 & 0.905 & 1.514 \\
E02 & 7 & 0.495 & 0.321 & 0.891 & 2.551 \\
E02 & 9 & 0.511 & 0.229 & 1.326 & 2.820 \\
E02 & 13 & 0.510 & 0.206 & 1.008 & 2.295 \\
E03 & 4 & 0.363 & 0.189 & 0.632 & 0.989 \\
E03 & 5 & 0.230 & 0.229 & 0.771 & 1.001 \\
E03 & 6 & 0.489 & 0.183 & 1.093 & 2.332 \\
E03 & 7 & 0.505 & 0.321 & 0.650 & 0.928 \\
E03 & 9 & 0.544 & 0.229 & 1.146 & 3.101 \\
E03 & 13 & 0.505 & 0.206 & 1.391 & 7.129 \\
\end{tabular}
\end{ruledtabular}
\end{table*}
\begin{table*}[t]
\caption{Historical job-associated calibration descriptors for the two faithful executions. Gate error is averaged over native two-qubit gates in each backward cone, weighted by their multiplicities; readout error is the maximum among the block\textquoteright s measured qubits. Percentages are calibration summaries, not fitted covariance attenuation or process fidelity. Placements, compilation, date and mitigation differ between runs.}
\label{sm:tab:sm_physical_historical}
\small
\begin{ruledtabular}
\begin{tabular}{lrrrrr}
Context & Size & $C_\alpha$ & Cone gates & Mean gate error (\%) & Max readout error (\%) \\
\hline
defaults & 4 & 0.864 & 14 & 0.210 & 0.537 \\
defaults & 5 & 0.183 & 144 & 0.209 & 1.636 \\
defaults & 6 & 0.143 & 46 & 0.202 & 0.708 \\
defaults & 7 & 0.479 & 55 & 0.164 & 1.367 \\
defaults & 9 & 0.571 & 158 & 0.213 & 1.245 \\
defaults & 13 & 0.670 & 162 & 0.236 & 1.733 \\
matched & 4 & 0.495 & 14 & 0.197 & 1.074 \\
matched & 5 & 0.234 & 144 & 0.220 & 1.270 \\
matched & 6 & 0.459 & 46 & 0.167 & 2.563 \\
matched & 7 & 0.506 & 55 & 0.202 & 1.343 \\
matched & 9 & 0.530 & 158 & 0.220 & 1.904 \\
matched & 13 & 0.501 & 162 & 0.194 & 1.929 \\
\end{tabular}
\end{ruledtabular}
\end{table*}

\section{IBM statistical estimands and uncertainty}
\label{sm:sm:statistics}
\subsection{Fixed-circuit three-job campaign}
\label{sm:sm:threecal}

The standardized campaign reuses, without re-transpilation, the archived family-$A$ joint
circuit, its 13 same-placement slices and the matched-mitigation faithful 51-qubit circuit.
Each of three epochs contains four contiguous rounds; within every round the same 15 circuits
are submitted in a newly seeded random order.  Every pub has $10^4$ shots, implemented as 20
twirling randomizations of 500 shots, with XY4 dynamical decoupling and gate and measurement
twirling.  The complete design therefore has 180 pubs and $1.8\times10^6$ shots.  IBM returns only the aggregate pub counts, so the 20
twirling realizations are not treated as independent strata.

For epoch $e$, round $r$ and logical block $i$, let $p^{J}_{eri}$ and $p^{S}_{eri}$ be the
probabilities that the joint pub and isolated slice, respectively, occupy one of that block's
ideal optimal configurations.  The first primary endpoint is
\begin{equation}
 R_{er}=\prod_{i\in\mathcal G}\frac{p^{J}_{eri}}{p^{S}_{eri}},
 \qquad
 R_e=\exp\!\left(\frac14\sum_{r=1}^{4}\log R_{er}\right),
 \label{sm:eq:sm_threecal_joint}
\end{equation}
where $\mathcal G$ comprises the six blocks containing two-qubit gates.  The second endpoint
uses the faithful circuit's measured energy $E_{er}$ and the energy $E_{{\rm prod},er}$ of the
product distribution with the same one-bit marginals,
\begin{equation}
 C_{er}=\frac{E_{{\rm prod},er}-E_{er}}
 {E_{\rm prod}^{\rm id}-E_{\rm id}},
 \qquad C_e=\frac14\sum_{r=1}^{4} C_{er}.
 \label{sm:eq:sm_threecal_faithful}
\end{equation}
Thus $R_e$ is a multiplicative joint/slice contrast and $C_e$ is the retained fraction of the
ideal weighted-covariance contribution.

We generate 4000 counting-model bootstrap draws per epoch from the archived aggregate
pub counts, using the archived order seed plus 991.  Multinomial draws preserve the
within-shot covariance of the joint-block indicators and of the faithful energy and
marginals; a binomial draw suffices for each separate slice's success indicator.
Each job-level draw averages one draw from each of its four \emph{fixed} observed rounds,
on the log scale for $R$ and identity scale for $C$.  Percentile intervals therefore
describe conditional counting uncertainty only.  They assume independent shots within
each pub and do not recover variation between the unavailable twirling strata.

Round variation is reported separately.  On the appropriate scale write
$y_{er}=\mu_e+a_{er}+\epsilon_{er}$ as a working additive model, with independent
round effects and counting errors of variance $v_{er,\mathrm{shot}}$.
The observed round variance already contains counting variation.  Its
method-of-moments excess is
$\widehat\tau^2_{e,\mathrm{round}}=\max(0,s^2_{e,\mathrm{round}}-\bar v_{e,\mathrm{shot}})$.
We do not resample noisy round estimates and then add shot draws, which would count
part of the counting variance twice.  Independence of rounds is a working assumption,
not established by the random pub order; temporal autocorrelation is not estimated.

For the third descriptive level, the archived properties-update timestamp associated with each completed job
labels the epoch; distinct timestamps do not imply statistically independent replicates. The
estimated variance of an epoch mean is
\begin{equation}
 \begin{aligned}
 \widehat v_e&=\frac{\bar v_{e,{\rm shot}}+
 \widehat\tau^2_{e,{\rm round}}}{4},\\
 \widehat\tau^2_{\rm job}&=\max\!\left(0,s^2_{\rm epoch}
 -\frac13\sum_{e=1}^{3}\widehat v_e\right).
 \end{aligned}
 \label{sm:eq:sm_threecal_variance}
\end{equation}
This subtraction is descriptive with three jobs; no stable interval for
$\widehat\tau^2_{\rm job}$ is claimed.  Under the working model, empirical dispersion
among job estimates already includes lower-level estimation variance.  Adding it again
to a $t$ interval would not repair the weak replication.  We report no job-population $t$ interval.

\begin{table}[t]
\caption{Fixed-circuit jobs, labeled by their associated properties snapshots (UTC$-3$). Brackets are conditional counting-model 95\% percentile intervals for the four-round job estimand, from 4000 bootstrap draws with the observed rounds held fixed. They do not quantify drift or between-job uncertainty.}
\label{sm:tab:sm_threecal}
\setlength{\tabcolsep}{3pt}
\begin{ruledtabular}
\begin{tabular}{lccc}
Epoch (snapshot) & Endpoint & Estimate & Conditional interval \\
\hline
E01 (Oct.~8, 08:55) & $R_e$ & 0.7931 & $[0.7716,0.8151]$ \\
 & $C_e$ & 0.5035 & $[0.4794,0.5279]$ \\
E02 (Oct.~14, 14:27) & $R_e$ & 0.9043 & $[0.8794,0.9285]$ \\
 & $C_e$ & 0.4997 & $[0.4758,0.5240]$ \\
E03 (Oct.~14, 18:25) & $R_e$ & 0.9123 & $[0.8871,0.9384]$ \\
 & $C_e$ & 0.5002 & $[0.4748,0.5241]$ \\
\end{tabular}
\end{ruledtabular}
\end{table}

For $R_e$, the geometric three-job mean is 0.8681.  The observed log-scale variance
$\RObservedVariance$ exceeds the mean epoch-estimation variance $\RMeanEstimationVariance$
by a factor $\RVarianceRatio$.  Equation~(\ref{sm:eq:sm_threecal_variance}) gives excess
$\RExcessVariance$, driven mainly by the lower E01 value.

For $C_e$, the arithmetic mean is 0.5011.  Its observed variance $\CObservedVariance$ is
only $\CVariancePercent\%$ of the mean epoch-estimation variance $\CMeanEstimationVariance$,
so the truncated excess is zero.  Thus the joint/slice contrast has positive descriptive
between-job excess, while the faithful correlation-retention endpoint has no resolved
excess.  With three jobs, two on the same day, neither result establishes a population
variance, long-term stationarity or a calibration field as the cause.

\textbf{Exploratory sensitivity.} Table~\ref{sm:tab:sm_sensitivity} omits each job in turn, recomputing observed dispersion and the mean working estimation variance for the remaining two jobs. Omitting E01 makes the log-$R$ ratio $0.19$ and the truncated excess zero; the other omissions retain positive excess. None resolves excess for $C$. Two-job summaries are omission diagnostics, not estimates of a job population.

We also declare hypothetical shot-variance multipliers $D\in\{1,2,5,10,20,50\}$ and reapply the same working subtraction with
\begin{equation}
 v_e(D)=\tfrac14\max\{s^2_{e,\rm round},D\bar v_{e,\rm shot}\}.
 \label{sm:eq:sm_inflation}
\end{equation}
The three-job log-$R$ excess reaches zero at $D=\RInflationThreshold$ within this sensitivity range; $C$ has zero excess throughout. This does not estimate shot dependence, validate independent rounds or reconstruct the unavailable twirling strata. It states the variance change required within this specified working model, not a significance level or calibrated robustness claim. Outputs are in \path{revision_v7/generated/campaign_sensitivity.json}.

\begin{table}[t]
\caption{Exploratory omission of one job. $s^2/\bar v$ compares observed dispersion with the mean working estimation variance, on the log scale for $R$ and identity scale for $C$. Each row has two jobs and is a sensitivity description, not population inference.}
\label{sm:tab:sm_sensitivity}
\begin{ruledtabular}
\begin{tabular}{lrrrr}
Omitted & Mean $R$ & $s_R^2/\bar v_R$ & Mean $C$ & $s_C^2/\bar v_C$ \\
\hline
E01 & 0.9083 & 0.19 & 0.5000 & 0.001 \\
E02 & 0.8506 & 49.65 & 0.5019 & 0.034 \\
E03 & 0.8469 & 40.50 & 0.5016 & 0.039 \\
\end{tabular}
\end{ruledtabular}
\end{table}

\subsection{Exploratory block covariance profiles}
\label{sm:sm:block_profiles}
The same faithful cost is evaluated in two historical executions with $10^5$ shots each and in the four fixed faithful-circuit rounds of E01--E03 with $10^4$ shots each. These 14 pubs contain $3.2\times10^5$ shots, not independent replicates of a device population. Historical point estimates motivated the exploration; the campaign is an additional observed context, not an external confirmatory validation.

For a support-graph block $V_\alpha$, define
\begin{equation}
 \Delta_\alpha=-\sum_{i<j\in V_\alpha}Q_{ij}\operatorname{Cov}(x_i,x_j),\quad
 C_\alpha=\frac{\Delta_\alpha}{\Delta_\alpha^{\rm id}}.
 \label{sm:eq:sm_block_covariance}
\end{equation}
The exact reference is reconstructed from the slack equalities and the executed angles. The six nontrivial block denominators sum to $29.8002036791\ldots$; the seven singletons have zero contribution and are not assigned a normalized ratio. With $w_\alpha=\Delta_\alpha^{\rm id}/\sum_\nu\Delta_\nu^{\rm id}$,
\begin{equation}
 C_{\rm global}=\sum_\alpha w_\alpha C_\alpha,\qquad
 G_\alpha=C_\alpha-C_{\rm global}.
 \label{sm:eq:sm_profile_contrast}
\end{equation}
The weights are positive in this reference, but that is not guaranteed for a general QUBO. Blocks of 9 and 13 variables carry $\BlockLargeWeight\%$ of the global denominator; the five-variable block, bases $\{0,1\}$ and slack bits $\{31,32,33\}$ in zero-based logical indexing, carries only $\BlockFiveWeight\%$.

\path{revision_v7/block_covariance.py} draws 10,000 multinomial replicates of each full 51-bit pub distribution. Uniform-category index occupancies are used as an exactly equivalent multinomial implementation when all observed count multiplicities coincide. All block moments in a replicate use the same bitstring weights, preserving their joint estimation covariance. Root seed 2026102102 creates 14 distinct streams with \texttt{SeedSequence.spawn} in the recorded source order. For each campaign job, statistics are first computed within each of its four fixed rounds and then averaged. Pooling first moments across rounds before forming covariances would change the estimand and is not used.

Pointwise percentile 95\% intervals are given in Table~\ref{sm:tab:sm_block_profiles}. The same draws provide energy $E$, product energy $E_{\rm prod}$, uniform-reference reductions $S=1-E/407$ and $S_{\rm prod}=1-E_{\rm prod}/407$, and their ratio. Job marginal shares are ratios of mean reductions, not averages of round ratios. No reduction near zero occurs here. The product share is approximately $90\%$ in all five contexts; it is a reference-distribution summary, not a generally bounded fraction or a quantum-performance certificate.

The single exploratory contrast family contains all six $G_\alpha$ in all five contexts, 30 entries in total. Let $\widehat s_j$ be a contrast's bootstrap standard deviation. The 95th percentile of
\begin{equation}
 T^*=\max_j\left|\frac{G_j^*-\widehat G_j}{\widehat s_j}\right|
 \label{sm:eq:sm_profile_maximum}
\end{equation}
gives simultaneous approximate bands $\widehat G_j\pm q_{0.95}\widehat s_j$. All estimated variances are nonzero; constant contrasts would be handled separately. Here $q_{0.95}=\BlockMaxQuantile$; split-half quantiles are 3.1298 and 3.1289 and give identical classifications. No covariance-matrix inversion or selection of extreme blocks is involved. Conditional independent-pub shot draws are a working model, not evidence that job effects or twirling strata are independent. These intervals omit unavailable stratum variability and temporal drift, as do the counting intervals above.

Table~\ref{sm:tab:sm_block_contrasts} shows the five-variable contrast below zero in all five contexts. In E01--E03 its ratios are $\BlockFiveEOne$, $\BlockFiveETwo$ and $\BlockFiveEThree$, with simultaneous contrast bands $[\BlockFiveEOneLow,\BlockFiveEOneHigh]$, $[\BlockFiveETwoLow,\BlockFiveETwoHigh]$ and $[\BlockFiveEThreeLow,\BlockFiveEThreeHigh]$. The observation concerns the block's weighted moment under the counting model. It does not determine individual edge covariances or a physical error mechanism. The two historical runs differ in placement, compilation, date and mitigation; their differences are not mitigation effects.

\begin{table*}[t]
\caption{Exploratory block weighted-covariance profiles. Brackets are pointwise conditional 95\% percentile intervals from 10,000 joint full-bitstring bootstrap draws per pub. Job entries average the four fixed round statistics. The seven singleton blocks have zero covariance contribution and no normalized ratio. Matched denotes the historical XY4-and-twirling execution; placements and dates differ from defaults.}
\label{sm:tab:sm_block_profiles}
\scriptsize
\setlength{\tabcolsep}{3pt}
\begin{ruledtabular}
\begin{tabular}{rrrccccc}
Size & $\Delta_\alpha^{\rm id}$ & $w_\alpha$ (\%) & Defaults & Matched & E01 & E02 & E03 \\
\hline
4 & 0.6277 & 2.11 & $0.864$ $[0.781,0.947]$ & $0.495$ $[0.393,0.596]$ & $0.307$ $[0.159,0.456]$ & $0.439$ $[0.291,0.585]$ & $0.363$ $[0.214,0.507]$ \\
5 & 1.1042 & 3.71 & $0.183$ $[0.119,0.247]$ & $0.234$ $[0.156,0.311]$ & $0.116$ $[-0.005,0.237]$ & $0.285$ $[0.163,0.403]$ & $0.230$ $[0.101,0.356]$ \\
6 & 2.8054 & 9.41 & $0.143$ $[0.098,0.187]$ & $0.459$ $[0.416,0.503]$ & $0.503$ $[0.432,0.574]$ & $0.520$ $[0.451,0.589]$ & $0.489$ $[0.419,0.560]$ \\
7 & 3.1008 & 10.41 & $0.479$ $[0.437,0.520]$ & $0.506$ $[0.462,0.550]$ & $0.502$ $[0.432,0.573]$ & $0.495$ $[0.424,0.564]$ & $0.505$ $[0.435,0.574]$ \\
9 & 7.7905 & 26.14 & $0.571$ $[0.543,0.599]$ & $0.530$ $[0.500,0.560]$ & $0.539$ $[0.493,0.585]$ & $0.511$ $[0.465,0.558]$ & $0.544$ $[0.497,0.589]$ \\
13 & 14.3716 & 48.23 & $0.670$ $[0.648,0.692]$ & $0.501$ $[0.478,0.525]$ & $0.523$ $[0.486,0.561]$ & $0.510$ $[0.472,0.547]$ & $0.505$ $[0.468,0.541]$ \\
Global & 29.8002 & 100 & $0.561$ $[0.546,0.576]$ & $0.495$ $[0.480,0.511]$ & $0.504$ $[0.479,0.528]$ & $0.500$ $[0.475,0.524]$ & $0.500$ $[0.476,0.525]$ \\
\end{tabular}
\end{ruledtabular}
\end{table*}

\begin{table*}[t]
\caption{Block-minus-global contrasts $G_\alpha=C_\alpha-C_{\rm global}$. Brackets are approximate simultaneous conditional 95\% bands for the single exploratory family of all 30 entries, using the standardized bootstrap maximum. An asterisk indicates exclusion of zero under the counting model, without a causal or population interpretation.}
\label{sm:tab:sm_block_contrasts}
\scriptsize
\begin{ruledtabular}
\begin{tabular}{rccccc}
Size & Defaults & Matched & E01 & E02 & E03 \\
\hline
4 & $0.303$ $[0.171,0.435]^{*}$ & $-0.001$ $[-0.160,0.159]^{}$ & $-0.197$ $[-0.434,0.040]^{}$ & $-0.061$ $[-0.293,0.172]^{}$ & $-0.138$ $[-0.371,0.095]^{}$ \\
5 & $-0.378$ $[-0.479,-0.276]^{*}$ & $-0.261$ $[-0.383,-0.140]^{*}$ & $-0.387$ $[-0.578,-0.197]^{*}$ & $-0.215$ $[-0.403,-0.028]^{*}$ & $-0.270$ $[-0.471,-0.069]^{*}$ \\
6 & $-0.418$ $[-0.487,-0.350]^{*}$ & $-0.036$ $[-0.104,0.031]^{}$ & $-0.001$ $[-0.111,0.109]^{}$ & $0.020$ $[-0.087,0.127]^{}$ & $-0.012$ $[-0.120,0.097]^{}$ \\
7 & $-0.082$ $[-0.147,-0.017]^{*}$ & $0.010$ $[-0.056,0.077]^{}$ & $-0.001$ $[-0.107,0.105]^{}$ & $-0.005$ $[-0.111,0.101]^{}$ & $0.005$ $[-0.101,0.110]^{}$ \\
9 & $0.010$ $[-0.029,0.050]^{}$ & $0.035$ $[-0.007,0.076]^{}$ & $0.035$ $[-0.030,0.100]^{}$ & $0.012$ $[-0.053,0.076]^{}$ & $0.043$ $[-0.021,0.108]^{}$ \\
13 & $0.110$ $[0.085,0.134]^{*}$ & $0.006$ $[-0.020,0.032]^{}$ & $0.020$ $[-0.021,0.060]^{}$ & $0.010$ $[-0.030,0.050]^{}$ & $0.004$ $[-0.036,0.044]^{}$ \\
\end{tabular}
\end{ruledtabular}
\end{table*}

\subsection{Historical executions and exploratory controls}
The QAOA sampler explores all $2^{24}$ strings.  Its primary concentration must therefore
compare the unconditional optimum probability with the unconditional null,
\begin{equation}
 \eta_{\rm all}=\frac{P(\mathrm{opt})}{288/2^{24}}.
\end{equation}
A conditional diagnostic is also legitimate provided that both numerator and denominator
are conditioned on $k=10$,
\begin{equation}
 \eta_{k=10}=\frac{P(\mathrm{opt}\mid k=10)}{288/\binom{24}{10}}.
\end{equation}
Dividing unconditional $P(\mathrm{opt})$ by the conditional denominator mixes sample spaces
and is not used.  Table~\ref{sm:tab:sm_stats} reports binomial Wilson intervals, which remain
well behaved at these small counts.

\begin{table*}[t]
 \caption{PEF-24 optimum statistics from $10^5$ shots per circuit, with the exact ideal
 values from the block factorization.  Confidence intervals are 95\% Wilson intervals
 propagated through the fixed null denominators.  The last two columns split the
 hardware-to-ideal probability ratio exactly into sector mass and within-sector selection,
 $\Phw(\mathrm{opt})/P_{\rm id}(\mathrm{opt})=[P_{\rm hw}(k{=}10)/P_{\rm id}(k{=}10)]
 \times[\eta_{k=10}/\eta_{k=10}^{\rm id}]$.  The $p=2,3$ angles are tiled from $p=1$.}
 \label{sm:tab:sm_stats}
 \setlength{\tabcolsep}{4pt}
 \begin{ruledtabular}
  \begin{tabular}{ccccccccccc}
   $p$ & Hits & $\eta_{\rm all}$ & 95\% CI & $\eta_{\rm all}^{\rm id}$
   & $P_{\rm hw}(k{=}10)$ & $P_{\rm id}(k{=}10)$ & $\eta_{k=10}$ & $\eta_{k=10}^{\rm id}$
   & sector ratio & within ratio \\
   \hline
   1 & 79 & 46.02 & $[36.93,57.34]$ & 106.34 & 0.1535 & 0.1747 & 35.05 & 71.17 & 0.879 & 0.492 \\
   2 & 48 & 27.96 & $[21.09,37.07]$ & 27.28 & 0.1256 & 0.1399 & 26.02 & 22.80 & 0.898 & 1.141 \\
   3 & 26 & 15.15 & $[10.34,22.19]$ & 9.92 & 0.1020 & 0.1117 & 17.36 & 10.38 & 0.913 & 1.673 \\
  \end{tabular}
 \end{ruledtabular}
\end{table*}

\textbf{Where the $p=1$ deficit sits.}  The uniform weight-ten mass is $0.1169$; the ideal
$p=1$ state raises it to $0.1747$ and the device to $0.1535$, so the device retains
$88\%$ of the ideal sector mass.  Inside the sector the ideal state selects a minimum cover
with probability $1.045\times10^{-2}$ ($\eta_{k=10}^{\rm id}=71.2$) and the device with
$5.15\times10^{-3}$ ($\eta_{k=10}=35.05$, Wilson interval $[28.14,43.64]$).  The
$0.433$ ratio is therefore $0.879\times0.492$, and $\ln0.492/\ln0.433=0.85$ of the
logarithmic deficit is intra-sector.  The device preserved cardinality and lost roughly half
of the selection among same-cardinality strings.  At $p=2,3$ the tiled-angle ideal state is
itself poorly concentrated and the within-sector ratio exceeds one; those rows are not a
decoherence series.

\textbf{Is the loss local to the blocks?}  The ideal state and the set of 288 optima are both
products over the 13 blocks, so for the ideal state
$P(\mathrm{opt})=\prod_iP_i(\mathrm{Opt}_i)$ exactly, where $P_i(\mathrm{Opt}_i)$ is the
probability that block $i$ is in one of its optimal configurations.  Evaluated on the measured
counts, the same product keeps everything the device does inside each block and removes every
correlation between blocks.  Table~\ref{sm:tab:sm_blocknull} compares it with the observed hits,
using a permutation test that shuffles each block's columns independently across shots
($2\times10^4$ permutations; \texttt{calc\_block\_product\_null.py}).  At $p=1$ the null
predicts $96.7\pm9.8$ hits against 79 observed ($p=0.07$, two-sided); at $p=2,3$ observation
and null agree.  No pair of block indicators has a correlation coefficient above $0.01$ in magnitude at any depth,
and independent blocks reproduce the measured weight-ten mass, so weight-preserving exchanges
between blocks are not detected either.  The $p=1$ ratio therefore splits as
$0.433=0.530\times0.817$: a within-block retention $\prod_iP^{\rm hw}_i/P^{\rm id}_i=0.530$
and a cross-block factor $0.817$ that is not resolved at $10^5$ shots.  The largest observed
within-block contribution is the four-qubit block, but block identity, qubit placement,
schedule and two-qubit-gate count are confounded.  The seven single-qubit blocks carry no two-qubit gate and retain
$0.978$--$1.013$ of their ideal probability; the two-qubit blocks retain $0.962$--$0.971$ and the
three-qubit blocks $0.901$--$0.995$; the four-qubit block retains $0.701$ and alone accounts for
$56\%$ of the within-block logarithmic loss.  This block cut and the
cardinality cut of the previous paragraph are two independent decompositions of the same
ratio, not successive steps:
\begin{equation*}
 0.433=\underbrace{0.879}_{\text{sector mass}}\times\underbrace{0.492}_{\text{selection in sector}}
      =\underbrace{0.530}_{\text{within blocks}}\times\underbrace{0.817}_{\text{between blocks}}.
\end{equation*}
The between-block factor has bootstrap 95\% interval $[0.65,1.00]$, compatible with no
cross-block deficit and with loss as large as $35\%$; no deficit is resolved.  Its equality
with the success proxy $\Fest=0.817$ is a coincidence.

\textbf{Placement and separately compiled blocks.} The principal circuit's 49 two-qubit gates remain in components $4,3,3,3,2,2$ and seven singletons. No gate touches an unmeasured qubit; different blocks have no directly coupled measured qubits. Layouts and instructions are archived. Job \texttt{dan62g5r85ps73fdlp00} ran 13 independently transpiled blocks with the same angles, XY4 and twirling, $10^4$ shots per block, reusing seven physical qubits. Their retention product is 0.784 against 0.530 in the joint run; the four-qubit block gives 0.932 against 0.701. Job-associated resource proxies are 0.822 and 0.817 and agree within 0.6\% per block, so they do not predict the observed contrast. Different day, placement and compilation remain confounded [main Fig.~\ref*{fig:blocks}].

\textbf{Joint and separate execution in one batch.} Job \texttt{dau04mahcrkc73dur170} (2026-09-29 associated snapshot, XY4 and twirling) ran compilation $A$ jointly with $10^5$ shots and then its 13 slices with $10^4$ shots each. A slice retains exactly the physical instructions on its block's qubits. Compilation $A$ has 47 two-qubit gates versus 49 for the principal compilation $B$; only one logical qubit has the same placement, although both four-qubit blocks use physical qubits 2,3,4,16. Table~\ref{sm:tab:sm_batch} gives the block results. The joint optimum count is 100, or 0.548 of the ideal probability. Within-block retention products are 0.542 jointly and 0.599 alone, ratio 0.906 $[0.848,0.965]$; the cross-block factor is 1.01 $[0.82,1.20]$. The gated-block product ratio is 0.886 $[0.849,0.925]$ and singleton ratio 1.022. Intervals condition on aggregate counts: the fixed joint-first order, unequal budgets and unmeasured drift remain limitations.

\textbf{Interleaved replication.} Job \texttt{daul9oqhcrkc73dvsqj0} (2026-09-30 snapshot, 368 s QPU time) resubmitted archived compilations $A$ and $B$ with their 13 slices and six separately compiled multi-qubit circuits $C$. Each of four rounds contains all 34 circuits in seeded random order, $10^4$ shots per pub in 20 twirls of 500 shots. No circuits are re-transpiled. Joint-block cones are disjoint and each slice retains its family's two-qubit instructions (Table~\ref{sm:tab:sm_interleaved}).

The gated-block geometric round ratios are 0.877 ($A$) and 0.671 ($B$). Conditional aggregate-count intervals are $[0.852,0.904]$ and $[0.650,0.692]$. The four-round log-scale $t$ summaries $[0.816,0.943]$ and $[0.662,0.681]$ describe realized-round dispersion rather than validated population uncertainty; counting variation is already present and individual twirling strata are unavailable. Pooled ratios including singletons are 0.869 and 0.691; singleton-only ratios are 0.990 and 1.029.

In $B$, block $\{0,1\}$ retains 0.898 alone and 0.640 jointly; its qubit 10 has $T_2=24~\upmu$s versus 71--142 $\upmu$s in $A$. Joint execution changes idle time and decoupling schedules, so the contrast includes these effects. The four-qubit retentions are 0.691/0.690 ($A$, joint/alone), 0.652/0.655 ($B$) and 0.929 ($C$), with resource proxies 0.949,0.948,0.931. Joint optimum counts are 37 and 27 in $4\times10^4$ shots, with cross-block factors 0.98 $[0.67,1.31]$ and 1.09 $[0.70,1.53]$. Gate lists, coherence times and counts are archived in the files listed in Sec.~\ref{sm:sm:data}; the analyses do not isolate a crosstalk mechanism.

\begin{table}[tbp]
 \caption{Joint and separate execution in one job on the same qubits
 (\texttt{dau04mahcrkc73dur170}).  Retention is $P_{\rm hw}(\mathrm{Opt}_i)/P_{\rm
 ideal}(\mathrm{Opt}_i)$; ``ratio'' is joint over separate, with a bootstrap 95\% interval.
 $\Fest$ is the job-time proxy of each block's gates and readout, identical for both
 executions.}
 \label{sm:tab:sm_batch}
 \setlength{\tabcolsep}{2.5pt}
 \scriptsize
 \begin{ruledtabular}
  \begin{tabular}{ccccccc}
   size & physical qubits & $n_{2Q}$ & $\Fest$ & joint & alone & ratio \\
   \hline
   4 & 2,3,4,16 & 18 & 0.946 & 0.699 & 0.709 & 0.986 $[0.963,1.009]$ \\
   3 & 140,141,142 & 7 & 0.971 & 0.934 & 0.962 & 0.971 $[0.959,0.984]$ \\
   3 & 109,110,118 & 9 & 0.968 & 0.933 & 0.952 & 0.980 $[0.969,0.992]$ \\
   3 & 20,21,36 & 9 & 0.972 & 0.978 & 0.993 & 0.985 $[0.958,1.014]$ \\
   2 & 90,91 & 2 & 0.988 & 0.952 & 0.978 & 0.974 $[0.969,0.979]$ \\
   2 & 105,117 & 2 & 0.988 & 0.959 & 0.975 & 0.984 $[0.975,0.993]$ \\
   1 & 42 & 0 & 0.997 & 0.994 & 0.988 & 1.006 $[0.986,1.027]$ \\
   1 & 53 & 0 & 0.996 & 1.017 & 1.009 & 1.008 $[0.988,1.028]$ \\
   1 & 128 & 0 & 0.995 & 1.003 & 0.997 & 1.007 $[0.987,1.027]$ \\
   1 & 59 & 0 & 0.995 & 1.014 & 1.003 & 1.011 $[0.992,1.032]$ \\
   1 & 60 & 0 & 0.997 & 0.981 & 0.985 & 0.996 $[0.984,1.009]$ \\
   1 & 149 & 0 & 0.996 & 0.991 & 1.000 & 0.991 $[0.975,1.007]$ \\
   1 & 77 & 0 & 0.994 & 0.996 & 0.993 & 1.002 $[0.986,1.020]$ \\
  \end{tabular}
 \end{ruledtabular}
\end{table}

\begin{table*}[tbp]
 \caption{Interleaved job \texttt{daul9oqhcrkc73dvsqj0}: within-block retention
 $P_{\rm hw}(\mathrm{Opt}_i)/P_{\rm ideal}(\mathrm{Opt}_i)$ of each block, pooled over four rounds
 ($4\times10^4$ shots per entry), for the two compilations $A$ and $B$ of the joint circuit, jointly
 (J) and as a light-cone slice alone (S), and for the separately transpiled block circuits $C$.}
 \label{sm:tab:sm_interleaved}
 \setlength{\tabcolsep}{2pt}
 \scriptsize
 \begin{ruledtabular}
  \begin{tabular}{ccccccccc}
   size & logical & $A$ qubits & $A$ J & $A$ S & $B$ qubits & $B$ J & $B$ S & $C$ S \\
   \hline
   4 & 2,3,4,5 & 2,3,4,16 & 0.691 & 0.690 & 2,3,4,16 & 0.652 & 0.655 & 0.929 \\
   3 & 6,7,9 & 140,141,142 & 0.926 & 0.971 & 89,90,91 & 0.925 & 0.940 & 0.963 \\
   3 & 12,13,14 & 109,110,118 & 0.911 & 0.936 & 38,49,50 & 0.918 & 0.928 & 0.943 \\
   3 & 15,16,17 & 20,21,36 & 0.986 & 0.990 & 148,149,150 & 1.005 & 1.006 & 1.018 \\
   2 & 0,1 & 90,91 & 0.945 & 0.982 & 10,11 & 0.640 & 0.898 & 0.976 \\
   2 & 21,22 & 105,117 & 0.957 & 0.973 & 109,110 & 0.952 & 0.978 & 0.980 \\
   1 & 8 & 42 & 1.003 & 1.005 & 86 & 1.012 & 1.000 & --- \\
   1 & 10 & 53 & 1.026 & 0.996 & 47 & 1.011 & 0.998 & --- \\
   1 & 11 & 128 & 1.019 & 1.000 & 77 & 1.005 & 0.995 & --- \\
   1 & 18 & 59 & 0.995 & 1.004 & 59 & 1.021 & 0.995 & --- \\
   1 & 19 & 60 & 0.984 & 0.994 & 128 & 0.983 & 0.997 & --- \\
   1 & 20 & 149 & 0.975 & 0.998 & 155 & 0.994 & 0.997 & --- \\
   1 & 23 & 77 & 0.997 & 1.009 & 124 & 0.980 & 0.995 & --- \\
   \hline
   \multicolumn{3}{l}{product} & 0.519 & 0.597 & & 0.340 & 0.493 & 0.820 \\
  \end{tabular}
 \end{ruledtabular}
\end{table*}

\begin{table}[tbp]
 \caption{Block-product null for the PEF-24 circuits.  ``Null'' is the mean and standard
 deviation of optimum hits over $2\times10^4$ independent per-block permutations; $p$ is
 two-sided.  $P(k{=}10)$ compares the measured weight-ten mass with the one implied by
 independent blocks.}
 \label{sm:tab:sm_blocknull}
 \setlength{\tabcolsep}{3pt}
 \begin{ruledtabular}
  \begin{tabular}{ccccccc}
   $p$ & hits & null & $p$ value & max $|r|$ & \shortstack{$P(k{=}10)$\\measured} & \shortstack{$P(k{=}10)$\\indep.\ blocks} \\
   \hline
   1 & 79 & $96.7\pm9.8$ & 0.073 & 0.010 & 0.1535 & 0.1524 \\
   2 & 48 & $43.3\pm6.5$ & 0.51 & 0.010 & 0.1256 & 0.1255 \\
   3 & 26 & $20.9\pm4.6$ & 0.32 & 0.007 & 0.1020 & 0.1026 \\
  \end{tabular}
 \end{ruledtabular}
\end{table}

For the faithful 51-qubit circuit, $53{,}569/10^5$ samples have base bits that form a cover,
with Wilson interval $[0.53260,0.53878]$.  Because the measured mean base cardinality is
$20.072$ rather than the optimum 10, the uniform 24-bit null is not a comparable reference.
Weighting the exact within-sector cover probabilities by the observed cardinality histogram
gives the matched null $0.447023$ and enhancement $1.19835$.  No measured 51-bit string
satisfies all slack equalities; the Wilson upper 95\% bound is $3.8413\times10^{-5}$.
Accordingly base-bit coverage is reported as a marginal diagnostic, not encoded feasibility.

\textbf{Energy nulls for the faithful encoding.}  Energy reduction relative to uniform bits
is dominated by the linear terms, so we compare the measured mean energy with two
correlation-free references evaluated exactly from the archived counts
(\texttt{calc\_hardware\_decomposition.py}).  The product distribution with the measured
single-bit marginals has mean energy $262.27$, a $35.6\%$ reduction and $90\%$ of the
measured $39.7\%$.  A null that keeps the measured base-cardinality histogram, draws bases
uniformly within each weight and keeps the slack marginals gives $239.45$, below the measured
$245.56$.  For the ideal $p=1$ state the product-marginal reference is $238.98$ against the
exact $209.18$, so $85\%$ of the ideal reduction is itself marginal.  The correlation energy
$E_{\rm prod}-E$ is $29.8$ ideally and $16.7$ measured: the device retains $56\%$ of it.  For
the 24-qubit $p=1$ circuit the same product null reaches only $43\%$ of the measured
reduction, so there the correlations carry most of the energy gain.

\textbf{Matched-mitigation repetition.}  The first faithful-encoding job used sampler
defaults while the QAOA jobs used XY4 decoupling and gate and measurement twirling.  We
therefore re-executed the identical logical circuit and angles with the QAOA settings (job
\texttt{dat665lvr3kc73el167g}; 372 two-qubit gates, depth 251).  Table~\ref{sm:tab:sm_exact51}
compares the two executions.  The matched run is not better: its energy reduction is
$36.6\%$ against $39.7\%$, its retained correlation energy $50\%$ against $56\%$, and its
job-time readout survival is lower ($0.652$ against $0.705$).  Neither execution satisfies
any complete set of slack equalities.  The base cover rate falls by $19\%$ relative
between the runs ($0.5357$ to $0.4326$), but the cardinality-matched nulls move in the
same direction ($0.4470$ against $0.3745$) and the matched enhancements nearly coincide
($1.198$ against $1.155$): the difference is dominated by the shift in base cardinality
($\langle k\rangle=20.07$ against $19.38$), not by a loss of cover structure.  Because the two
jobs used different transpilations, placements, dates and properties snapshots, they do not
isolate a mitigation effect.  The correlation-energy diagnostic persists under both settings.

\begin{table}[tbp]
 \caption{The two executions of the faithful 51-qubit encoding ($10^5$ shots each).
 $E_{\rm prod}$ is the product-marginal null; ``corr.\ retained'' is
 $(E_{\rm prod}-E)/(E_{\rm prod}^{\rm id}-E^{\rm id})$ with ideal correlation energy $29.8$.}
 \label{sm:tab:sm_exact51}
 \setlength{\tabcolsep}{3pt}
 \scriptsize
 \begin{ruledtabular}
  \begin{tabular}{lccc}
   & defaults & XY4+twirling & ideal \\
   \hline
   job & \texttt{dakjjpfi\ldots} & \texttt{dat665lv\ldots} & --- \\
   $\Fest$ & 0.316 & 0.306 & --- \\
   $\langle E\rangle$ (uniform 407.0) & 245.56 & 258.07 & 209.18 \\
   energy reduction & 39.7\% & 36.6\% & 48.6\% \\
   fraction of ideal gap & 0.82 & 0.75 & 1 \\
   $E_{\rm prod}$ & 262.27 & 272.82 & 238.98 \\
   product-null share of reduction & 0.90 & 0.90 & 0.85 \\
   corr.\ retained & 0.56 & 0.50 & 1 \\
   $\langle k\rangle$ bases & 20.07 & 19.38 & --- \\
   base cover rate & 0.5357 & 0.4326 & --- \\
   card.-matched null & 0.4470 & 0.3745 & --- \\
   all slack equalities & 0 & 0 & --- \\
  \end{tabular}
 \end{ruledtabular}
\end{table}

\textbf{Routing and correlation between logical blocks.}  The faithful operator is block-local
over 13 logical blocks, and its ideal state is a product across them, so every correlation
between block observables vanishes ideally.  The executed circuit is not block-local on the
chip.  Hardware noise can correlate two blocks through gates only if their backward light cones
overlap, that is, if some instruction can influence the measured qubits of both.  We compute
the cones from the archived transpiled circuits (\texttt{ibm\_circuits/exact51\_*.ops.json},
retrieved by \texttt{fetch\_ibm\_circuits.py}), following every routing SWAP, rather than from
the logical-to-physical map at measurement.  The two executions are separate transpilations
(only four of the 51 logical qubits are measured on the same physical qubit), yet in both exactly
three of the 78 pairs have overlapping cones: the blocks of 13, 9 and 5 variables, which share
135 two-qubit gates (72 of them lie in all three cones) on 27 physical qubits.  All other cones
are disjoint, and in the 24-qubit circuit all 13 are.  We computed the correlation between the
per-shot energies of every pair of logical blocks from the archived counts
(\texttt{calc\_exact51\_cross\_block.py}).  The two most correlated pairs of the 78 are, in both
executions, the pairs 13--9 and 9--5.  We do not assign a rank probability to that observation,
because the 78 pairs differ in block size and variance and are not exchangeable.  Their correlations are
$z=r\sqrt{N_s}=4.0$ and $3.7$ with sampler defaults and $3.3$ and $6.2$ with matched mitigation, with
permutation $p$ values of at most $9\times10^{-4}$ ($10^4$ permutations); three of the four exceed
the Bonferroni threshold $3.41$ for all 78 pairs, and the fourth ($3.3$) is no larger than the
largest value of the 24-qubit contrast group.  The third overlapping pair, 13--5, is not
correlated ($|z|\le0.6$, permutation $p\ge0.57$).  The 75 pairs with disjoint cones stay within
$|z|\le2.6$.  One of them has measured qubits adjacent on the chip and stays within $|z|\le0.3$,
which gives no evidence of readout crosstalk on that coupler but does not test it elsewhere
[Fig.~\ref*{fig:exact51}(c)].  The effect is small, $r=0.010$--$0.020$, and the summed
cross-block energy covariance is about $0.5\%$ of the energy variance. For a block-additive cost, dependence between blocks at fixed block marginals does not change the mean energy; cross-block energy covariances contribute to its estimation variance. The observed contribution is small in these runs. In these two exploratory transpilations, correlations detectable after the stated pairwise tests are
confined to blocks whose light cones overlap, but one overlapping pair is not correlated.  In the 24-qubit circuit
the same statistic stays below the Bonferroni threshold for all 78 pairs (largest $|z|=3.2$;
Table~\ref{sm:tab:sm_blocknull}).  An earlier classification by the map at measurement, which
labels a pair routed when a gate joins the qubits that carry the two blocks at readout, finds
36 such gates and misses the pair 13--5; it is kept in \texttt{exact51\_cross\_block.json} for
comparison.

\section{Mixer-control results}
\subsection{Same-sector mixer control}
\label{sm:sm:same_sector_overview}
The quantum walk optimization algorithm (QWOA) \cite{marsh2020combinatorial,hadfield2019quantum,cook2020quantum} replaces the transverse-field mixer with a cardinality-preserving $XY$ mixer on a Dicke state.  A raw comparison with QAOA therefore changes the mixer, input and accessible sector together: handing QWOA $k=k^\star$ supplies information unavailable to full-register QAOA.  The raw median $P(\mathrm{opt})$ ratio is $7.1\times10^4$, while QAOA places median probability $1.68\times10^{-4}$ on that sector.  The ratio is primarily a cardinality prior, not an isolated mixer effect.

The $p=1$ convergence audit uses twenty task seeds on one fixed cluster of $\nB=12$ posts, with 23 targets per seed, strict $m=1,2,3$, and radii selected in 500--720 km. Its supports contain 49--66 of the 66 possible edges. The generator and its alternative radius rule are specified in Sec.~\ref{sm:sm:details}.  Both treatments minimize the same exact one-sided objective,
\begin{equation}
 E(x)=|x|+(\nB+1)\sum_t \mathbf 1\!\left[(Ax)_t<m\right],
 \label{sm:eq:qwoa_audit_cost}
\end{equation}
and $P(\mathrm{opt})$ is evaluated only after the angles are fixed.  The instance, $k^\star$ sector, Dicke input, optimizer and depth are held fixed; only the RMS-normalized shared-task edge weights are changed to unit weights on the same support.

The weighted-to-control median $P(\mathrm{opt})$ ratio is $1.02$, with seed-level bootstrap interval $[0.95,1.17]$; seed-aggregated Wilcoxon tests give $p=0.26$ for probability and $p=0.16$ for energy.  Thus no benefit is detected, but effects from a $5\%$ loss to a $17\%$ gain remain compatible with the data.  The evolution is evaluated spectrally: no digital QWOA circuit was compiled or executed, and no $p>1$ point met the prespecified convergence rule.  An exploratory unit-weight support-versus-complete comparison also finds no detectable difference (seed-level $p_E=0.28$, $p_P=0.50$); four supports are already complete. These tests concern task variation on this dense, fixed geography. Raw-weight and pilot-exclusion controls are reported in the Supplemental Material.

\subsection{Twenty-seed same-sector comparison}
\label{sm:sm:qwoa20}
The pilot's six rows came from two task seeds, and the seed-to-seed dispersion was large
($P_Q(k^\star)$ differs by a factor 240 between the two seeds at $m=1$).  We therefore
repeated the complete $p=1$ protocol on twenty distinct seeds drawn without replacement from a freshly initialized generator,
\path{default_rng(20260915).choice(24,20,replace=False)}, which contains seeds 12 and 17
(\texttt{run\_qwoa\_20seeds.py}, \texttt{qwoa\_20seeds.json}).  Each of the 60 (seed, $m$)
rows optimizes five ans{\"a}tze with the protocol of Sec.~\ref{sm:sm:settings}, adaptive third
resolution included: QAOA, and QWOA with the raw weighted, RMS-normalized weighted,
support-matched unit-weight and complete unit-weight mixers.  The six pilot endpoints are
reproduced exactly.  QAOA and every normalized or unit-weight endpoint pass the joint
convergence rule in all 60 rows; the raw weighted mixer passes in 51, and its comparisons use
those rows only.  The scale-equivalence criterion of Table~\ref{sm:tab:sm_qwoa_mixer_control},
$\beta^\star_{\rm raw}\mathrm{RMS}\le\pi$, holds in 49 of 60 rows.

\textbf{Cardinality prior.}  QWOA's unconditional $P(\mathrm{opt})$ exceeds QAOA's in all 60
rows (median ratio $7.1\times10^4$, range $3.3$--$1.6\times10^7$), while QAOA places a median
$1.68\times10^{-4}$ (range $7.3\times10^{-6}$--$0.072$) on the supplied sector.  Conditioning
QAOA on $k^\star$ lowers the median ratio to $2.8$ (range $0.16$--$4.8\times10^3$; QWOA larger in
43 of 60).  QAOA has the lower expected energy in all 60 rows, which the sector restriction largely
explains (Sec.~\ref{sm:sm:qwoa_convergence}).

\textbf{Mixer effect.}  Table~\ref{sm:tab:sm_qwoa20} gives the four same-sector contrasts.  Row
counts treat the three multiplicities of a seed as separate observations, which they are not:
they share the instance and its cooperation graph.  The inferential unit is therefore the
seed, and we report seed-level paired tests on the per-seed mean energy difference and mean
log probability ratio.  At row level two contrasts look suggestive: the normalized weighted
mixer has lower energy in 39 of 60 rows (sign test $p=0.027$), and the raw weighted mixer has
larger $P(\mathrm{opt})$ than the complete graph in 36 of 51 ($p=0.005$).  Neither survives
aggregation by seed, where every contrast has $p\ge0.15$.  The median probability ratio of the
primary contrast is $1.02$ with interquartile range $0.91$--$1.11$, and it is the same at every
multiplicity (medians $1.03$, $1.01$, $1.03$ at $m=1,2,3$).  Resampling seeds gives a bootstrap
95\% interval $[0.95,1.17]$ for the seed-level median.  Within single multiplicities, the raw
weights against the support control give row-level Wilcoxon $p=0.012$ (energy) and $0.018$
(probability) at $m=1$; over the family of twelve per-$m$ tests (two contrasts, three
multiplicities, two metrics) the Holm-adjusted values are all $\ge0.15$.

\textbf{Sensitivity to the pilot seeds.}  Seeds 12 and 17 fixed the protocol, so we repeat the
primary seed-level contrast without them.  Over the remaining 18 seeds the normalized weighted
mixer has lower energy in 11 and larger $P(\mathrm{opt})$ in 11 (Wilcoxon $p=0.18$ for both), with
median probability ratio $1.02$ and bootstrap interval $[0.95,1.17]$.

The primary same-support comparison detects no change attributable to shared-task weights at $p=1$ on this fixed-site family. A separate exploratory topology contrast is given below.  This is absence of evidence at the stated precision, not a demonstration of
equivalence: no equivalence margin was fixed in advance, and the interval admits effects from
a $5\%$ loss to a $17\%$ gain.

\begin{table*}[t]
  \caption{Twenty-seed same-sector $p=1$ mixer contrasts ($\nB=12$, $m=1,2,3$).  ``$A$ wins''
  counts rows (left block) or seeds (right block) in which the treatment $A$ has the lower
  expected energy or the larger $P(\mathrm{opt})$.  Seed-level tests use each seed's mean
  energy difference and mean log $P(\mathrm{opt})$ ratio over its converged rows; $p$ values
  are two-sided Wilcoxon signed-rank.  $W_{\rm RMS}$: RMS-normalized shared-task weights;
  $W_{\rm raw}$: raw weights; $U$: unit weights on the same support; $K$: complete unit-weight
  graph.  The row count $n$ is 60 for contrasts with $W_{\rm RMS}$ and 51 for contrasts with
  $W_{\rm raw}$, because the raw weighted mixer meets the joint convergence rule in only 51
  rows and comparisons use converged endpoints only; every other ansatz converges in all 60.}
  \label{sm:tab:sm_qwoa20}
  \setlength{\tabcolsep}{4pt}
  \scriptsize
  \begin{ruledtabular}
    \begin{tabular}{lcccccc|cccc}
      & \multicolumn{6}{c|}{rows} & \multicolumn{4}{c}{seeds (20)} \\
      contrast $A$ vs $B$ & $n$ & $E$ wins & $P$ wins & median $P_A/P_B$ & IQR & median $\Delta E$
      & $E$ wins & $P$ wins & $p_E$ & $p_P$ \\
      \hline
      $W_{\rm RMS}$ vs $U$ (primary) & 60 & 39 & 36 & 1.021 & 0.91--1.11 & $-0.087$ & 12 & 12 & 0.16 & 0.26 \\
      $W_{\rm raw}$ vs $U$ & 51 & 33 & 32 & 1.036 & 0.91--1.21 & $-0.085$ & 9 & 11 & 0.15 & 0.37 \\
      $W_{\rm RMS}$ vs $K$ & 60 & 34 & 36 & 1.025 & 0.89--1.16 & $-0.085$ & 12 & 13 & 0.57 & 0.26 \\
      $W_{\rm raw}$ vs $K$ & 51 & 29 & 36 & 1.080 & 0.95--1.29 & $-0.063$ & 10 & 13 & 0.50 & 0.16 \\
    \end{tabular}
  \end{ruledtabular}
\end{table*}

\textbf{Exploratory topology-only comparison.} To change only support, compare unit weights on the cooperation support ($U$) with unit weights on the complete graph ($K$), using their archived converged endpoints. Each seed contributes the mean energy difference and mean log probability ratio over $m=1,2,3$. Four seeds have complete supports and exactly zero differences. The two-sided Wilcoxon tests use \texttt{zero\_method=wilcox}, the normal approximation and no continuity correction. Median seed effects have pointwise percentile intervals from 10,000 seed-bootstrap draws with seed 20261021. The two endpoint tests are reported raw and with Holm adjustment within this exploratory family only; they do not acquire the status of the primary contrast. Table~\ref{sm:tab:sm_topology} gives a median probability ratio one, interval $[0.976,1.066]$, and raw $p_E=0.28$, $p_P=0.50$. The dense, fixed-site supports limit the scope of this negative observation. The analysis in \path{revision_v7/mixer_summary.py} performs no new optimization.

\begin{table}[t]
\caption{Exploratory topology-only contrast $U-K$ from the 20 archived seeds. Effects are seed-level medians; probability is $P_U/P_K$. Brackets are pointwise 95\% seed-bootstrap intervals (10,000 draws). Signed-rank tests discard exact zeros and use the normal approximation. Holm adjustment covers these two endpoints only.}
\label{sm:tab:sm_topology}
\begin{ruledtabular}
\begin{tabular}{lrrr}
Metric & Effect [interval] & $p$ & $p_{\rm Holm}$ \\
\hline
Energy & $0.000$ $[-0.073,0.340]$ & 0.28 & 0.56 \\
Probability & $1.000$ $[0.976,1.066]$ & 0.50 & 0.56 \\
\end{tabular}
\end{ruledtabular}
\end{table}

\subsection{Convergence-first two-seed pilot}
\label{sm:sm:qwoa_convergence}
Table~\ref{sm:tab:sm_qwoa_convergence} reports every converged $p=1$ pilot point.  Lower
expected energy and larger optimum probability are different questions here: QAOA places
most of its mass on feasible but nonminimum high-cardinality strings, whereas QWOA is given
the minimum-cardinality sector.  QAOA's lower expected energy in all six cases is therefore
largely explained by the sector restriction rather than by the mixers.  It is not forced:
a QWOA state concentrated on the optima would reach energy $k^\star$, below every QAOA
endpoint.  But with $k^\star$ small, most strings of the supplied sector are infeasible, and at $m=1$, $k^\star=2$ a single violation already costs
$2+13=15$, whereas QAOA reaches $7.6$ on feasible high-cardinality strings of the full
register.  The same-sector control of Sec.~\ref{sm:sm:qwoa_mixer_control}, where every
ansatz lives in the same sector, brings all energies into the common range $25$--$31$.
QWOA has larger unconditional optimum probability in all six cases.  The latter ratio
is inflated by the small QAOA sector mass shown explicitly in the table.  Conditioning on
$k^\star$ reverses both $m=1$ comparisons, but still does not create a same-sector training
control because the QAOA angles minimize the full-register expectation.

\begin{table*}[t]
  \caption{Converged $p=1$ expectation-value audit.  $P_Q(k^\star)$ is QAOA's probability
  of the sector supplied to QWOA; $P_Q(\mathrm{opt}\mid k^\star)$ uses the same QAOA state
  conditioned on that sector.  The last column is
  $P_W(\mathrm{opt})/P_Q(\mathrm{opt}\mid k^\star)$.  It is a cardinality diagnostic, not
  a controlled mixer effect.  The energy comparison is driven largely by the sector
  restriction (see text) and is not a mixer effect.}
  \label{sm:tab:sm_qwoa_convergence}
  \setlength{\tabcolsep}{4pt}
  \scriptsize
  \begin{ruledtabular}
    \begin{tabular}{ccccccccc}
      seed & $m$ & $k^\star$ & $\langle E\rangle_W$ & $\langle E\rangle_Q$
      & $P_W(\mathrm{opt})$ & $P_Q(\mathrm{opt})$ & $P_Q(k^\star)$
      & conditional ratio \\
      \hline
      12 & 1 & 2 & 27.281 & 7.631 & 0.2247 & $8.07\!\times\!10^{-3}$ & $2.26\!\times\!10^{-2}$ & 0.629 \\
      12 & 2 & 5 & 26.873 & 11.485 & 0.3213 & $1.92\!\times\!10^{-6}$ & $1.38\!\times\!10^{-4}$ & 22.96 \\
      12 & 3 & 8 & 26.307 & 12.166 & 0.2774 & $3.98\!\times\!10^{-7}$ & $1.83\!\times\!10^{-4}$ & 127.65 \\
      17 & 1 & 2 & 29.526 & 8.280 & 0.0966 & $1.07\!\times\!10^{-5}$ & $9.55\!\times\!10^{-5}$ & 0.863 \\
      17 & 2 & 5 & 28.385 & 11.498 & 0.0584 & $8.29\!\times\!10^{-7}$ & $1.81\!\times\!10^{-4}$ & 12.76 \\
      17 & 3 & 8 & 28.731 & 12.184 & 0.0515 & $4.96\!\times\!10^{-8}$ & $1.64\!\times\!10^{-4}$ & 170.46 \\
    \end{tabular}
  \end{ruledtabular}
\end{table*}

The two-resolution global search initially disagreed for QWOA at $(12,2)$ and $(17,3)$.
The adaptive 4096-point/three-iteration run reproduced the lower-energy endpoint in the
first case and the coarse endpoint in the second to better than $10^{-8}$ in both energy
and $P(\mathrm{opt})$, providing two-resolution replication for every row.  At $p=2$ and
3, by contrast, zero of six pairs met the convergence rule.  We therefore report no depth
trend and do not compute inferential ensemble statistics from their best-known values.

\subsection{Pilot scale and topology controls}
\label{sm:sm:qwoa_mixer_control}
The cardinality confound was removed by placing both ans{\"a}tze in the same $k^\star$
sector, starting both in the same uniform Dicke state and using the same exact cost and
$p=1$ convergence protocol.  The primary control preserves the cooperation-graph support
but replaces its shared-task weights by one.  Thus only the edge weights change.  The raw
weighted mixer and this support-matched control each have lower expected energy in three
of six cases.  The raw weighted mixer has larger $P(\mathrm{opt})$ in four cases, but its
median ratio is only 1.10 (range 0.483--1.98).

The common finite $\beta\in[0,\pi]$ window is not invariant under a global rescaling of
the mixer.  Defining the root mean square as
$\operatorname{RMS}(w)=[|E|^{-1}\sum_{e\in E}w_e^2]^{1/2}$, the shared-task weights have
RMS 9.63 for seed 12 and 10.36 for seed 17, whereas unit weights have RMS one.
Table~\ref{sm:tab:sm_qwoa_mixer_control} therefore reports the scale-controlled sensitivity
analysis $w_{bb'}\mapsto w_{bb'}/\operatorname{RMS}(w)$.  All
endpoints meet the same joint energy-and-probability replication criterion as
Table~\ref{sm:tab:sm_qwoa_convergence}.  Weighted and uniform mixers each win three energy
comparisons and three probability comparisons; the median probability ratio is 0.975
(range 0.300--1.98).  No inferential test is attached to these six rows, which come from
only two independent task seeds; the twenty-seed extension of Sec.~\ref{sm:sm:qwoa20} supplies
one.

\begin{table*}[t]
  \caption{Converged same-sector $p=1$ mixer control on the two pilot seeds.  $W_{\rm raw}$
  is the shared-task-weighted mixer, $W_{\rm RMS}$ the same mixer divided by its edge-weight
  RMS, and $U$ the support-matched unit-weight control.  All use the same $k^\star$ sector,
  Dicke state, exact cost, optimizer and window $\beta\in[0,\pi]$.  Since
  $e^{-i\beta H_{XY}(w/\mathrm{RMS})}=e^{-i(\beta/\mathrm{RMS})H_{XY}(w)}$, the raw optimum
  $\beta^\star_{\rm raw}$ is reachable by the normalized mixer iff
  $\beta^\star_{\rm raw}\,\mathrm{RMS}\le\pi$.  This holds in four rows, where the two
  endpoints coincide exactly and $\beta^\star_{\rm RMS}=\beta^\star_{\rm raw}\,\mathrm{RMS}$.
  It fails for seed 17 at $m=1,2$ ($\beta^\star_{\rm raw}\,\mathrm{RMS}=29.5$ and $21.6$),
  where the raw mixer had exploited an effective evolution time ten times longer than the
  window allows the normalized one, and the normalized optimum is a different endpoint.}
  \label{sm:tab:sm_qwoa_mixer_control}
  \setlength{\tabcolsep}{3.5pt}
  \scriptsize
  \begin{ruledtabular}
    \begin{tabular}{ccc|cccc|cccc|ccc|cc}
      & & & \multicolumn{4}{c|}{$W_{\rm raw}$} & \multicolumn{4}{c|}{$W_{\rm RMS}$}
      & \multicolumn{3}{c|}{$U$} & & \\
      seed & $m$ & $k^\star$
      & $\langle E\rangle$ & $P(\mathrm{opt})$ & $\beta^\star$ & $\beta^\star\mathrm{RMS}$
      & $\langle E\rangle$ & $P(\mathrm{opt})$ & $\beta^\star$ & same?
      & $\langle E\rangle$ & $P(\mathrm{opt})$ & $\beta^\star$
      & $\frac{P_{W_{\rm raw}}}{P_U}$ & $\frac{P_{W_{\rm RMS}}}{P_U}$ \\
      \hline
      12 & 1 & 2 & 27.281 & 0.2247 & 0.051 & 0.49 & 27.281 & 0.2247 & 0.491 & yes & 25.573 & 0.1936 & 1.761 & 1.160 & 1.160 \\
      12 & 2 & 5 & 26.873 & 0.3213 & 0.011 & 0.10 & 26.873 & 0.3213 & 0.103 & yes & 31.135 & 0.1620 & 0.049 & 1.983 & 1.983 \\
      12 & 3 & 8 & 26.307 & 0.2774 & 0.010 & 0.10 & 26.307 & 0.2774 & 0.100 & yes & 26.775 & 0.2665 & 3.072 & 1.041 & 1.041 \\
      17 & 1 & 2 & 29.526 & 0.0966 & 2.845 & 29.5 & 29.725 & 0.0713 & 0.102 & no & 29.482 & 0.0784 & 0.070 & 1.233 & 0.909 \\
      17 & 2 & 5 & 28.385 & 0.0584 & 2.085 & 21.6 & 28.605 & 0.0363 & 0.275 & no & 29.401 & 0.1208 & 0.074 & 0.483 & 0.300 \\
      17 & 3 & 8 & 28.731 & 0.0515 & 0.011 & 0.11 & 28.731 & 0.0515 & 0.110 & yes & 26.733 & 0.0620 & 0.106 & 0.831 & 0.831 \\
    \end{tabular}
  \end{ruledtabular}
\end{table*}

As a secondary topology control, all pairs were coupled with unit weight.  Seed 12 already
has the complete 66-edge support, so that control is algebraically identical to $U$; seed
17 has 62 of 66 edges and was optimized independently.  Against the RMS-normalized weighted
mixer, the complete control has lower energy in four of six cases, while the
$P(\mathrm{opt})$ comparison again splits three--three.  Hence neither the relative weights
nor the four missing edges identify a stable advantage.  The generic control gradients
match central differences to $2.3\times10^{-9}$ or better.

\section{Six hundred instances on official base sites}
\label{sm:sm:ssec:ensemble}
\label{sm:ssec:ensemble}
\begin{figure*}[t]
  \centering
  \includegraphics[width=\linewidth]{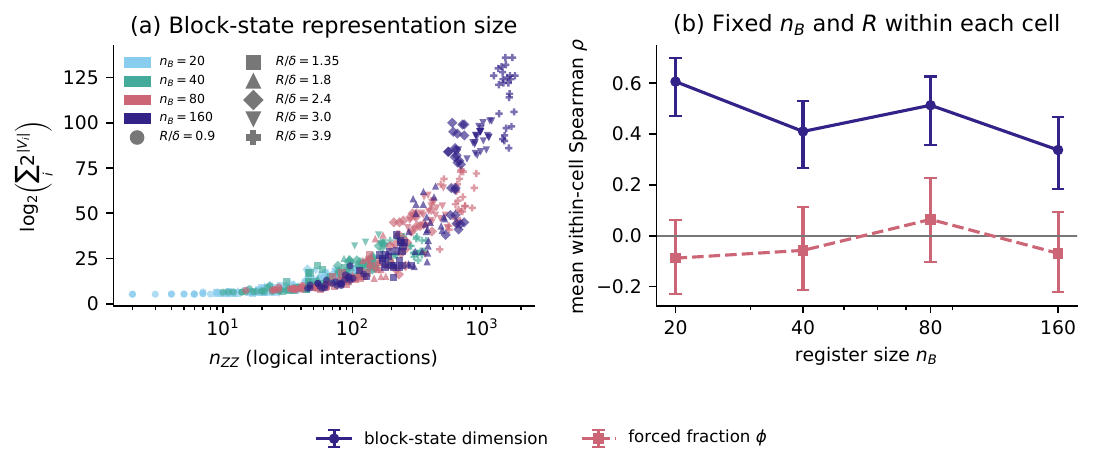}
  \caption{Structural descriptors over 600 feasible instances with certified optima. (a) Combined block-state dimension $D_{\rm block}$ versus logical interaction count; neither axis is a measured runtime. (b) Mean within-cell Spearman correlations with block-state dimension and forced fraction $\phi=n_f/\kopt$ (bases forced by tight constraints). Each summary averages six coefficients at each $\nB$, with radius fixed within each cell. Bars are pointwise 95\% percentile intervals from 4000 paired-instance bootstrap draws within each cell of 25 instances; the six radii are not resampled. Positions and incidence matrices vary across instances.}

  \label{sm:fig:sm_squeeze_ens}
\end{figure*}

To test block size beyond one frozen instance, we generated 600 feasible instances whose bases are sampled from the 773 official municipal centroids and whose synthetic targets are anchored near those bases, using four register sizes and six radii. Here $\delta$ is the median nearest-neighbor spacing of a fixed reference sample of $\nB$ municipalities (seed 0), not of each generated instance.  All 600 optima are certified by integer programming.  At $\nB=160$ the mean largest-block size grows from 12 variables at $R/\delta=0.9$ to 116 at $R/\delta=3.9$. Pooled over all sizes and radii, interaction count and block-state dimension have rank correlation $+0.97$.  The latter retains an instance-level signal after conditioning on register size and radius [Fig.~\ref{sm:fig:sm_squeeze_ens}].  The forced fraction $\phi=n_f/\kopt$ counts bases required by initially tight covering constraints, relative to the optimum size; it is a partial structural descriptor. This is an empirical result for the stated generator, not a universal relation between gate count and simulability.

The ensemble has four base counts $\nB\in\{20,40,80,160\}$, six fixed radius ratios
$R/\delta\in\{0.90,1.35,1.80,2.40,3.00,3.90\}$ and 25 realizations per cell.
For each cell let $f=(R/\delta)/3$; replicate $s=0,\ldots,24$ uses seed
$10000\nB+100\,\mathrm{int}(100f)+s$.  Bases are sampled without replacement from the
773 official centroids.  Each of $\nB$ initial targets has a uniformly chosen anchor
base, a uniform bearing and a uniform radial offset in $[0,0.7R]$.
Any target outside all great-circle coverage disks is dropped.  There is no greedy-size
acceptance filter, unlike the selected PEF-24 instance, but the ensemble remains
conditioned on reachable synthetic targets.

All 600 constrained ILPs are certified optimal within the 30-s solver cap.
This is an empirical result for this family, not a complexity classification.
The raw per-instance results are in \path{results/block_decomposition.json}
(\texttt{ensemble}) and \path{results/ensemble_squeeze.json} (\texttt{rows}).
They are joined by $(\nB,R/\delta,\mathrm{seed})$ for the revised analysis.

Within each of the 24 fixed cells we compute Spearman's correlation between
$\log\nzz$ and $\log_2 D_{\rm block}$, and between $\log\nzz$ and $\phi$.
For each $\nB$, the summary is the arithmetic mean of its six cell coefficients.
Each of 4000 bootstrap replicates resamples the 25 \emph{paired instances}
within each cell and recomputes that mean; the six radius settings are not resampled.
A draw with a constant variable is redrawn within its cell and counted in the analysis
record.  The seed is 20261016.  Table~\ref{sm:tab:sm_ensemble_correlations} gives pointwise percentile
intervals, not simultaneous confidence bands or a causal test.

\begin{table}[t]
\caption{Mean of the six within-cell Spearman coefficients at each register size. Brackets are pointwise 95\% percentile intervals from 4000 paired-instance bootstrap draws within each fixed cell; radii are not resampled. Each cell contains 25 instances.}
\label{sm:tab:sm_ensemble_correlations}
\begin{ruledtabular}
\begin{tabular}{ccc}
$n_B$ & Block-state dimension & Forced fraction $\phi$ \\
\hline
20 & $0.60$ $[0.47,0.70]$ & $-0.09$ $[-0.23,0.06]$ \\
40 & $0.41$ $[0.26,0.53]$ & $-0.06$ $[-0.21,0.11]$ \\
80 & $0.51$ $[0.36,0.62]$ & $0.06$ $[-0.11,0.23]$ \\
160 & $0.34$ $[0.18,0.47]$ & $-0.07$ $[-0.22,0.09]$ \\
\end{tabular}
\end{ruledtabular}
\end{table}

The average coefficients over all cells are $+0.47$ for block dimension and $-0.04$
for $\phi$.  Every size-specific block-dimension interval is positive, while each
$\phi$ interval includes zero.  This supports an instance-level association beyond
the shared radius setting in the sampled family; it neither makes $D_{\rm block}$
an exact runtime nor proves absence of a forced-fraction association elsewhere.
The separate frozen-target comparison is in Sec.~\ref{sm:sm:frozen}.

\section{Frozen-target control}
\label{sm:sm:frozen}
In the ensemble of Sec.~\ref{sm:sm:ssec:ensemble} each target is drawn within $0.7R$ of its anchor base, so
the target distribution dilates with $R$.  Table~\ref{sm:tab:sm_frozen} repeats the sweep with
the targets drawn once at the smallest radius and then held fixed, which is the design used
for the PEF-24 sweep.  At $\nB=160$ the two means differ by at most $0.04$ in $\phi$ at the four tested radii.
This particular target-dilation control does not remove the observed floor; it does not
exclude other generator effects.  With 20 bases and frozen targets, $\phi$ falls to $0.09$.

\begin{table}[tbp]
  \caption{Mean $\phi$ with frozen targets (15 instances per cell), against the anchored
    ensemble of Sec.~\ref{sm:sm:ssec:ensemble}.}
  \label{sm:tab:sm_frozen}
  \begin{ruledtabular}
    \begin{tabular}{ccccc}
      $\nB$ & $R/\delta$ & frozen & anchored & $\Delta$ \\
      \hline
      20  & 0.90 & 0.782 & 0.755 & 0.027 \\
      20  & 1.80 & 0.422 & 0.498 & 0.076 \\
      20  & 3.00 & 0.142 & 0.275 & 0.133 \\
      20  & 3.90 & 0.086 & 0.166 & 0.081 \\
      \hline
      160 & 0.90 & 0.808 & 0.810 & 0.002 \\
      160 & 1.80 & 0.519 & 0.558 & 0.039 \\
      160 & 3.00 & 0.283 & 0.320 & 0.037 \\
      160 & 3.90 & 0.191 & 0.205 & 0.014 \\
    \end{tabular}
  \end{ruledtabular}
\end{table}

\section{The forced fraction along the range dial}
\label{sm:sm:ssec:triviality}
\label{sm:ssec:triviality}
Table~\ref{sm:tab:squeeze} follows one frozen PEF-24 target set as $R$ increases.
Its forced fraction $\phi$ and connected components characterize the covering constraints;
$\nzz$ counts logical interactions.  From 200 to 600 km, $\nzz$ rises from 17 to 121
and the block-state dimension from 62 to $2^{24}$.  These are properties of the same
support graph, not a theorem that every inexpensive circuit must be easy to simulate.
A sparse graph can be connected, and a connected circuit can admit other efficient
classical representations.

In particular, $D_{\rm block}=\sum_i2^{|V_i|}$ measures the representation used by our
block simulator, not a simulation lower bound or a solver runtime.  Here the constrained
coverage optimum is certified in under a millisecond at every radius (median of 25
runs on the recorded machine).  Fragmentation is determined by shared targets, not
distance alone: 31 pairs in different blocks lie within $2R$, the closest 35 km apart.
The range sweep links interaction count to this particular representation; it does
not demonstrate classical hardness at its connected end.

\begin{table*}[tbp]
  \caption{Frozen PEF-24 range sweep.  State dimension is
  $D_{\rm block}=\sum_i2^{|V_i|}$, the dimension of a blockwise state-vector representation,
  not a runtime or lower bound.  The constrained ILP times are machine-dependent medians
  of 25 certified solves, archived in \texttt{dial\_sweep.json}.
  $\Fest$ uses the device-median proxy of Eq.~(\ref*{eq:resource}) with the PEF-24
  routing overhead; these hypothetical radii were not transpiled.  Consequently the
  200-km value 0.77 differs from the job-associated physical-gate proxy 0.82.}
\label{sm:tab:squeeze}
  \setlength{\tabcolsep}{2.5pt}
  \scriptsize
  \begin{ruledtabular}
    \begin{tabular}{cccccccccc}
      $R$ (km) & $\nzz$ & $\Fest$ & $\kopt$ & $n_f$ & $\phi$ & blocks & largest & state dim. & ILP (ms) \\
      \hline
      200 &  17 & 0.77 & 10 & 4 & 0.40 & 13 &  4 & $\mathbf{62}$ & $<1$ \\
      250 &  20 & 0.75 &  9 & 3 & 0.33 & 11 &  6 & $110$ & $<1$ \\
      300 &  29 & 0.71 &  8 & 2 & 0.25 &  9 &  7 & $174$ & $<1$ \\
      350 &  49 & 0.64 &  8 & 1 & 0.13 &  4 & 12 & $4{,}194$ & $<1$ \\
      400 &  65 & 0.58 &  6 & 1 & 0.17 &  3 & 17 & $131{,}138$ & $<1$ \\
      450 &  87 & 0.51 &  6 & 1 & 0.17 &  2 & 23 & $8{,}388{,}610$ & $<1$ \\
      500 &  98 & 0.48 &  4 & 0 & 0.00 &  1 & 24 & $16{,}777{,}216$ & $<1$ \\
      550 & 110 & 0.45 &  4 & 0 & 0.00 &  1 & 24 & $16{,}777{,}216$ & $<1$ \\
      600 & 121 & 0.42 &  3 & 0 & 0.00 &  1 & 24 & $\mathbf{16{,}777{,}216}$ & $<1$ \\
    \end{tabular}
  \end{ruledtabular}
\end{table*}

\section{Why blockade is not a native dominating-set encoding}
\label{sm:sm:ssec:rydberg}
\label{sm:ssec:rydberg}
Neutral-atom arrays evolve
\begin{equation}
H_{\mathrm{Ryd}}=\frac{\Omega}{2}\sum_i\sigma_x^i
-\sum_i\Delta_i\hat n_i+\sum_{i<j}\frac{C_6}{r_{ij}^6}\hat n_i\hat n_j
\end{equation}
\cite{jaksch2000fast,wurtz2023aquila}.  In the hard-blockade limit the positive pair term
forbids simultaneous excitations on adjacent vertices.  Its feasible bitstrings are therefore
independent sets.  A dominating set obeys a different predicate,
$\sum_{b\in N[t]}x_b\ge1$ for every target $t$; selected vertices may be adjacent and every
unselected vertex must have a selected neighbor.  Equality of the blockade graph and the
problem graph cannot make these predicates equivalent.

We audited the proposed $3\times3$ grid by exact enumeration.  The grid has domination
number three and ten minimum dominating sets.  With
$C_6=5.420441\times10^6\,\mathrm{rad}\,\upmu\mathrm{s}^{-1}\,\mu\mathrm{m}^6$ and
$\Omega=15\,\mathrm{rad}\,\upmu\mathrm{s}^{-1}$, the nominal blockade radius is
$R_b=8.44\,\upmu$m.  At $d=7\,\upmu$m the blockade graph does match the grid edges, yet only
eight of the ten minimum dominating sets are blockade-compatible; the sets
$\{3,4,5\}$ and $\{1,4,7\}$ are valid optima that blockade forbids.  At $d=5\,\upmu$m the
blockade also reaches diagonals and only four of the ten optima remain compatible.

For reproducibility, the statevector protocol starts in $|0\rangle^{\otimes9}$ and uses
$\Omega_{\max}=15\,\mathrm{rad}\,\upmu\mathrm{s}^{-1}$.  The Rabi frequency ramps linearly from
zero to $\Omega_{\max}$ over $0.3\,\upmu$s, remains flat for $\tau-0.6\,\upmu$s and ramps back to
zero over the final $0.3\,\upmu$s.  The global detuning is linear from
$-10$ to $+1\,\mathrm{rad}\,\upmu\mathrm{s}^{-1}$ over the full pulse.  Static local detunings are
$\Delta_i^{\rm loc}=2|N[i]|-1\,\mathrm{rad}\,\upmu\mathrm{s}^{-1}$ for the grid closed-neighborhood
sizes.  The archived trajectory uses $\tau=2.6\,\upmu$s and DOP853 with relative tolerance
$10^{-9}$ and absolute tolerance $10^{-11}$.

The final classical Hamiltonian supplies the decisive test.  At $d=7\,\upmu$m its unique
ground configuration is $\{0,2,6,8\}$, the four corners, which does not dominate the center.
At $d=5\,\upmu$m the two ground configurations are $\{0,8\}$ and $\{2,6\}$; neither dominates
the grid.  Thus the ground manifold contains zero minimum dominating sets and zero dominating
sets at both spacings.

At $\tau=2.6\,\upmu$s, the externally scored minimum-dominating-set probabilities are 0.075406 and 0.312737 at 5 and 7 $\upmu$m. At 7 $\upmu$m the actual ground probability is 0.003870 and blockade-independent-set probability 0.997995. These are transient overlaps with an externally defined target set, not evidence of a faithful MDS encoding. Furthermore, the bare quadratic surrogate weights $2\lambda|N[b]\cap N[b']|$ are generally incompatible with a distance-only interaction $C_6/r_{bb'}^6$; correcting those weights alone would still leave the equality-versus-covering issue. Explicit reductions using ancilla-atom gadgets can encode general QUBOs \cite{nguyen2023quantum}, but no such reduction is constructed here. This audit concerns only the stated direct vertex-to-atom mapping.

\bibliography{refs_v8}
\end{document}